\documentclass[aps,prd,showpacs,preprintnumbers]{revtex4}
\usepackage{graphics}
\def \beq{\begin{equation}}
\def \eeq{\end{equation}}
\def \beqa{\begin{eqnarray}}
\def \eeqa{\end{eqnarray}}
\def\etc{{\sl etc.\/}}
\def\ie{{\sl i.e.\/}}
\def \A{{\cal A}}
\def \B{{\cal B}}
\def \D{\widetilde D}

\def \M{{\mathbf M}}
\def \rey{\mathrm{Re}}
\def \re{{\bf Re}\,}
\def \S{\mathbf S}
\def \U{{\cal U}}
\def \V{\mathbf V}
\def \vis{\eta_{\scriptscriptstyle V}}
\def\etal{{\sl et al.\/}}
\def\annp{{\sl Ann.\ Phys.\/}}

\def\jhep{{\sl J.\ H.\ E.\ P.\/}}
\def\npa{{\sl Nucl.\ Phys.\/}, A}
\def\npb{{\sl Nucl.\ Phys.\/}, B}
\def\plb{{\sl Phys.\ Lett.\/}, B}

\def\prd{{\sl Phys.\ Rev.\/}, D}
\def\pre{{\sl Phys.\ Rev.\/}, E}
\def\progtp{{\sl Prog.\ Theor.\ Phys.\/}}

\begin{document}
\title{Aspects of causal viscous hydrodynamics}
\author{R.\ S.\ \surname{Bhalerao}}
\email{bhalerao@theory.tifr.res.in}
\affiliation{Department of Theoretical Physics, Tata Institute of Fundamental
         Research,\\ Homi Bhabha Road, Mumbai 400005, India.}
\author{Sourendu \surname{Gupta}}
\email{sgupta@tifr.res.in}
\affiliation{Department of Theoretical Physics, Tata Institute of Fundamental
         Research,\\ Homi Bhabha Road, Mumbai 400005, India.}

\begin{abstract}
\end{abstract}
\preprint{TIFR/TH/07-11}

\begin{abstract}
We investigate the phenomenology of freely expanding fluids, with
different material properties, evolving through the Israel-Stewart (IS)
causal viscous hydrodynamics, and compare our results
with those obtained in the relativistic Eckart-Landau-Navier-Stokes (ELNS) acausal viscous
hydrodynamics. Through the analysis of scaling invariants
we give a definition of thermalization time which can be self-consistently
determined in viscous hydrodynamics. Next we construct the solutions for
one-dimensional boost-invariant flows. Expansion of viscous fluids is
slower than that of one-dimensional ideal fluids, resulting in entropy
production. At late times, these flows are reasonably well approximated
by solutions obtained in ELNS
hydrodynamics.  Estimates of initial energy densities from observed
final values are strongly dependent on the dynamics one chooses. For
the same material, and the same final state, IS hydrodynamics gives
the smallest initial energy density.  We also study fluctuations about
these one-dimensional boost-invariant backgrounds; they are damped in
ELNS hydrodynamics but can become sound waves in IS hydrodynamics.
The difference is obvious in power spectra due to clear signals of
wave-interference in IS hydrodynamics, which is completely absent in
ELNS dynamics.
\end{abstract}
\maketitle

\section{Introduction}

Hydrodynamics is an effective long-distance theory of fluids close to
thermal and chemical equilibrium. The hydrodynamic modes are the fields
which enter the theory, and can be identified from the symmetries of the
microscopic theory. They include the energy-momentum tensor and densities
of conserved quantities such as the baryon number and electric charge.
The theory contains several unknown quantities, the transport coefficients
and relaxation times, which one must obtain from microscopic computations
in kinetic theory \cite{transport}, or through measurements.

The relativistic version of the Navier-Stokes equation was first explored
by Eckart \cite{eckart} and subsequently by Landau \cite{landau}, who
developed what we call the ELNS formalism. The relativistic equations
for ideal fluids are widely used in the contexts of heavy-ion collisions
\cite{bjorken,baym,kagiyama,idealhic} and cosmology \cite{idealcosmo}.
It has been suggested that the fluid produced in heavy-ion collisions at
RHIC is very close to ideal. Such a claim must be substantiated by a study
of viscous fluid dynamics. Until now there have been very few studies of
non-ideal fluids in this context \cite{nonidealhic,prakash,muronga,rischke,
heinz,baier,shuryak,romrom,dumitru}.

ELNS theory for non-ideal fluids is known to violate causality
\cite{acausal}. The problem can be traced to the linear relation between
fluxes and thermodynamic forces which is inherent in the Chapman-Enskog
method of obtaining the ELNS equations from kinetic theory.  It was
realized by Israel and Stewart \cite{causal} that the problem with
causality could be repaired by simply going beyond the linear relation
between fluxes and forces. Their formulation of hydrodynamics is
variously known as second-order theory, causal viscous hydrodynamics,
or Israel-Stewart (IS) hydrodynamics.  It contains an expanded set of
material parameters when compared to ELNS theory.

There are other attempts to repair the loss of causality in Navier-Stokes
theory. It was shown that the introduction of a phenomenologically
motivated lag between the application of a thermodynamic force and
the material response, through a memory kernel, could restore causality
\cite{koide}. Such a phenomenological approach contains a smaller
number of material parameters than IS hydrodynamics. In any case, little
is known about some of the new parameters which appear in IS
hydrodynamics. Even the relation between current correlation functions
in a thermal quantum field theory and these quantities \cite{postkubo}
has not been studied comprehensively.

In fact, once the problems of principle were resolved, further
investigations of causal viscous hydrodynamics lagged because of an
apparent paucity of applications. One expects that the main applications
would be in situations where either the mean free path is comparable to
the size of the region of interest or the relaxation time approaches
the time scale of the phenomena of interest. It turns out that such
applications are not hard to come by today. The former are possibly
realized in heavy-ion collisions \cite{rischke,heinz,baier},  and the
diffusion of neutrinos through supernovae, the latter in astrophysical
shock waves and freezeout in relativistic reactive fluids. Interest in
such systems is on the increase.

Very little is presently known about the nature of fluid flows in
IS hydrodynamics. The present paper is a step towards repairing this
neglect, keeping future applications to heavy-ion physics in mind. In
common with \cite{muronga,rischke,heinz,baier} we investigate the equations for a fluid
characterized completely by the energy-momentum tensor, \ie, neglecting
net baryon number and electric charge, keeping only the shear part
of the viscous stress tensor. We set up the equations in curvilinear
coordinates appropriate to the approximate boost-invariant geometry of
heavy-ion collisions, and reduce the tensor equations to coupled scalar
equations. By examining the symmetries of the hydrodynamic equations we
obtain laws of physical similarity.

Since an ideal fluid has zero viscosity and zero mean free path, the ideal
hydrodynamic equations are unable to predict their own failure. All
estimates of thermalization and freezeout in ideal hydrodynamics
are imposed from outside.  Since causal viscous hydrodynamics contains the
relaxation time, $\tau_\pi$, one expects to improve upon this.
We present a preliminary analysis.

Next we analyze boost-invariant solutions with three sets of
constitutive relations for the fluid. In each case, we compare the
ELNS and IS descriptions of boost-invariant flow and find that the
former approximates the latter at late times. Our results for the case
of the massless Boltzmann fluid are consistent with those given in
\cite{muronga,rischke,baier}. We also examine the propagation
of fluctuations around the boost-invariant
solutions.  Here the differences between ELNS and IS descriptions are
remarkable: no propagating solutions exist in ELNS dynamics, whereas IS
dynamics gives rise to damped sound waves.

The plan of the paper is as follows.  The next section introduces
the hydrodynamic equations and extracts scaling laws from them. This
section also contains a discussion of the properties of materials that
are needed in the remainder of the paper. Following this, we present the
well-known Bjorken solution, and illustrate our method with the analysis
of fluctuations around this solution for the ideal fluid. The next
three sections examine a simple fluid, a Boltzmann fluid, and conformal
fluids, respectively.  In each case we examine the boost-invariant
solution in IS hydrodynamics, its approach to ELNS hydrodynamics at
late times, entropy production, and fluctuations around the
boost-invariant solution. Through this analysis we build up a picture of
general properties of the flow, as opposed to those which
are specific to certain kinds of fluids. We summarize our understanding in
the final section. The appendices contain details of the tensor analysis,
the reduction of the tensor hydrodynamic equations to coupled scalar
equations, and an analysis of transients.

\section{The hydrodynamic equations}

Throughout this paper we investigate the hydrodynamic equations in the
limit of zero net quarks, \ie, vanishing baryon and charge density,
since this is a good approximation to the actual situation realized
in ultra-relativistic heavy-ion collisions. 
We also ignore the heat flux as in \cite{muronga,rischke,heinz,baier}. 
References \cite{muronga20071,muronga20072} extend this to the case where
the heat flux, thermal conductivity and baryon density are all
included.
It is strongly suspected
that in the high-temperature phase of QCD, not very close to the
crossover temperature $T_c$, the bulk viscosity is negligible. In
view of this we investigate the equations where the bulk viscosity
has been set to zero. With these simplifications the explicit form of
the equations of Israel-Stewart (IS) hydrodynamics was written down in
\cite{muronga,heinz,baier}.

Having set the baryon and charge densities to zero, one is left with three
independent hydrodynamic variables: a scalar, a vector and a tensor. The
scalar is the energy density, $\epsilon$. It is related to the pressure,
$p$, by the equation of state--- $p=c_s^2\epsilon$, where $c_s$ is the
speed of sound. Since all three quantities in the equation of state can
be written as a function of the temperature, $T$, we sometimes trade
$\epsilon$ for $T$. This variable specifies the part of the stress tensor
from which external work can be extracted. Another of the hydrodynamic
variables is the velocity 4-vector $u^\mu$ (various choices of $u$
are discussed in \cite{landau} and \cite{causal}).  Finally one has the
dissipative part of the stress tensor. When bulk viscosity is neglected,
this is a traceless symmetric tensor, $\pi^{\mu\nu}$. Since this part
expresses shear, it must vanish on contraction with $u^\mu$.

Since we will discuss longitudinal flow, it is convenient to
go from Galilean coordinates $(t,x,y,z)$ to the curvilinear
coordinates $(\tau,\eta,r,\phi)$ where $\tau=\sqrt{t^2-z^2}$,
$\eta=\tanh^{-1}(z/t)$, $r=\sqrt{x^2+y^2}$ and $\phi=\tan^{-1}(y/x)$. The
metric becomes $g_{\mu\nu}=\mathrm{diag}(1,-\tau^2,-1,-r^2)$. The only
non-vanishing Christoffel symbols are $\Gamma^\tau_{\eta\eta}=\tau$,
$\Gamma^r_{\phi\phi}=-r$, $\Gamma^\eta_{\tau\eta}=\Gamma^\eta_{\eta\tau}=
1/\tau$ and $\Gamma^\phi_{r\phi}=\Gamma^\phi_{\phi r}=1/r$ \cite{ctf}. We will
write covariant derivatives as $d_\mu$ and partial derivatives as
$\partial_\mu$. The action of $d_\mu$ on a scalar field is the same
as the action of $\partial_\mu$. In terms of the proper time ${\cal
T}=\sqrt{\tau^2-r^2}$, one defines the components of the velocity 4-vector
$u^\mu=dx^\mu/d{\cal T}$. One can show that $u^\mu u_\mu=1$.

Longitudinal flow is an approximation applicable to ultra-relativistic
heavy-ion collisions when the hydrodynamic variables at any point of
spacetime depend on $\tau$ and $\eta$ but not on $r$ and $\phi$. Clearly
such an approximation is valid far from the edges of the fluid volume, and
at times $\tau\ll R/c_s$, where $R$ is a typical transverse size. We will
parametrize the velocity vector by a quantity $y$ in the form---
\beq
   u^\mu = (\cosh y,\frac1\tau\sinh y,0,0).
\label{udef}\eeq
We note that the rapidity is $y+\eta$.  Scaling flow corresponds to $y=0$
in our notation, as we discuss later.  A fluid element with $y=0$, in
our notation, moves along a world line of fixed $\eta$, corresponding
to a constant velocity $v=\tanh\eta$.  The divergence of $u$ is
\beq
   \Theta = d_\mu u^\mu = y_\tau\sinh y + (y_\eta+1)\frac1\tau\cosh y.
\label{thetadef}\eeq
This defines a macroscopic time scale for a hydrodynamic flow.
Here, and later, we use the notation $f_\tau$ \etc, to denote the derivative
of a scalar $f$ with respect to the variable $\tau$ {\etc}
We also define the material derivative $D=u^\mu d_\mu$, and through it
the spacelike vector $D u^\mu$ and its norm $S^2=-Du^\mu Du_\mu$. A straightforward
computation shows
\beq
   S = y_\tau\cosh y + (y_\eta+1)\frac1\tau\sinh y.
\label{sdef}\eeq
Then it is easy to write down the unit space-like vector $v^\mu=(Du^\mu)/S$,
\beq
   v^\mu = (\sinh y,\frac1\tau\cosh y,0,0).
\label{vdef}\eeq
One also defines another directional derivative operator $\D= v^\mu
d_\mu$. In the local rest frame one finds that $D$ is the derivative with
respect to time and $\D$ is the longitudinal spatial derivative.

Using the methods outlined in the appendices, one finds
the hydrodynamic equations---
\beqa
\nonumber
    D\epsilon+B\Theta\epsilon &=& \Theta\pi_V,\\
\nonumber
    c_s^2\D\epsilon + BS\epsilon &=& \D\pi_V + S\pi_V,\\
    \tau_\pi D\pi_V + \pi_V &=& \frac43 \vis\Theta,
\label{hydro}\eeqa
where $B=1+c_s^2$, $\vis$ is the coefficient of shear viscosity and
$\tau_\pi$ is the relaxation time associated with the shear stress,
$\pi_V$. Recall that such a relaxation time is necessary to construct
causal hydrodynamics \cite{causal,acausal}. For longitudinal flow we have
reduced the tensor equations of hydrodynamics to three coupled scalar
equations for the three scalar hydrodynamic quantities $\epsilon$, $y$
and $\pi_V$.

ELNS hydrodynamics is recovered when $\tau_\pi=0$, so that the last of
eqs.\ (\ref{hydro}) reduces to $\pi_V = 4\vis\Theta/3$. One expects that
for fluids which evolve inertially, \ie, in the absence of external forces
acting during the evolution, the solutions of eqs.\ (\ref{hydro}) should
approach the solutions of ELNS hydrodynamics at times $\tau\gg\tau_\pi$.
This conclusion may clearly change when a fluid is acted upon by external
forces at all times. In such cases, of course, driving terms have to be
added to the equations.

\subsection{Material properties at vanishing chemical potential}

For an ideal fluid, the equation of state can be cast into the form $p =
c_s^2\epsilon$, where $p$, $\epsilon$ and $c_s$, could all be functions
of the temperature, $T$.  Straightforward dimensional analysis shows
that $\epsilon = bT^4$, where $b$ is dimensionless. In general there
are various intrinsic mass scales, $\mu_i$, in the fluid, and $b$ could
have an implicit dependence on $T$ through the functional dependence,
$b(T/\mu_1,T/\mu_2,\cdots)$. If $c_s^2=1/3$ at all temperatures, then the
trace of the stress tensor vanishes identically. This implies a special
symmetry called scale symmetry, or conformal symmetry \cite{jackiw}. One
aspect of conformal symmetry is that uniform scaling of external scales
such as $T$ by a constant leaves material properties unchanged. Clearly,
then $b$ cannot depend on $T$, and must be constant.

For a non-ideal fluid, the stress tensor has an additional viscous
part. When the trace of the full stress tensor vanishes, then the bulk
viscosity vanishes identically.
Conformal symmetry implies that the physics of such
fluids can be expressed in terms of dimensionless combinations of
material properties which are temperature independent. One such
combination which has been used in the literature is $\vis/s$, where
$\vis$ is the coefficient of shear viscosity and $s=(\epsilon+p)/T$
is the entropy density. In passing we note that for a conformal fluid,
$s=\gamma\epsilon^{3/4}$, where $\gamma$ is a dimensionless constant.

Causal viscous hydrodynamics requires another material property
of the fluid, the relaxation time for the shear part of the viscous
stress tensor, $\tau_\pi$. Dimensionally, $\tau_\pi=a/T$, where $a$ is
dimensionless, and becomes constant when the fluid has conformal symmetry.
This dimensionless number is proportional to the quantity called liquidity
\cite{iitk} which, in non-relativistic fluids, measures the mean-free
path (proportional to $\tau_\pi$) in units of the interparticle spacing
(proportional to $1/s^{1/3}\sim1/T$). In a gas, this number is very
large, in liquids, small.  It follows from the expressions for $s$,
$\tau_\pi$ and the equation of state, that $s={\cal K}\epsilon\tau_\pi$,
where the dimensionless constant ${\cal K}=4/3a$. This implies that
the dimensionless material property $\epsilon\tau_\pi/\vis=s/{\cal
K}\vis$. We call this combination $\chi$ and discuss it extensively
in the next subsection.

In reality, the fluids that we are interested in are not conformal
\cite{bielefeld}.  At temperatures below $T_c$ the fluid of hadron
resonances has a plethora of mass scales, which breaks conformal symmetry,
and manifests itself in deviations of $c_s^2$ from the value $1/3$. It
is not a big stretch of the imagination to expect that bulk viscosity
will be non-vanishing in this fluid. A fluid of quarks and gluons also
breaks conformal symmetry through the conformal anomaly, which results in
the running of the strong coupling and the appearance of the QCD scale
$\Lambda_{QCD}$, and explicitly through the quark masses.  One question
of interest is how important are these departures from conformal symmetry.

One might expect that at very large $T$, when the QCD coupling is
close to zero, and all the quark masses are much less than $T$, one
might have conformal symmetry to a good approximation, by virtue of
the fluid being well-approximated by a massless ideal gas. In fact,
this is the limit in which the Boltzmann fluid approximation is seen to
hold in weak coupling theory, with $\chi=3/2c_s^2=9/2$. In this limit,
one has, additionally, bulk viscosity much smaller than shear viscosity
\cite{bulk}, $c_s^2\approx1/3$, and the energy density close to an ideal
gas value.

Lattice computations show, surprisingly, that approximate conformal
symmetry is obtained also at substantially smaller $T/T_c\approx$2--3,
where the pressure deviates significantly from its ideal gas
value \cite{swagato}. Toy models of QCD with substantially enhanced
($N=4$ super-) symmetries, which give up the running of the coupling
and asymptotic freedom, have been used to model this observation.
They are bound to fail in the vicinity of $T_c$ where the conformal
measure \cite{swagato} is large, and bulk viscosity cannot be neglected
\cite{dima}.  One prediction from these toy models, using the AdS/CFT
conjecture, is that $\vis/s=1/4\pi$, yielding $\chi=4\pi/{\cal
K}$. A recent computation in an appropriate $N=4$ SYM theory has
found $a=(1-\ln2)/6\pi$ \cite{hellerjanik}, which then yields
$\chi=(1-\ln2)/2\approx0.15$.

In this paper we shall examine three models of viscous fluids. The first,
which we call a simple fluid, is one in which the material properties
$c_s$, $\vis$ and $\tau_\pi$ are constant. Lattice results show that
$c_s$ is almost constant over a range of $T$ \cite{swagato}. However,
preliminary lattice computations of transport coefficients are
almost consistent with the power counting in $T$ over the same range
\cite{sigma}.  As a result, the main motivation to study this model
of a simple fluid is not its direct application to heavy-ion physics,
but the fact that it allows explicit computation of the hydrodynamics,
and contains qualitatively all the phenomena that we find with other
models of viscous fluids, as we show in a later section.

The more restricted models of materials that we use have the property
that $\chi$ is constant. A Boltzmann fluid has been examined in
the literature \cite{baier}, and is defined by the specific value
$\chi=9/2$. We devote one section to detailed hydrodynamics of the
Boltzmann fluid. In addition, in a subsequent section, we examine the
whole class of conformal fluids with $\chi=3\pi a$, for various $a$.
Note that the hydrodynamics of the conformal fluid
with $a=3/2\pi$ (\ie, $\tau_\pi=3/2\pi T$) is exactly equivalent to that
for a Boltzmann fluid. As a result, it does not seem possible to use
hydrodynamics alone to distinguish a conformal fluid from a Boltzmann
fluid. We discuss this in greater detail later.

\subsection{Laws of physical similarity}

The equations of ideal hydrodynamics
are obtained by setting $\pi_V=\vis=0$ in the eqs.\ (\ref{hydro}).
Then the equations for the remaining
hydrodynamic variables, $\epsilon$ and $y$, are---
\beq
    D\epsilon+B\Theta\epsilon = 0,\qquad
    c_s^2\D\epsilon + BS\epsilon = 0.
\label{ideal}\eeq
Consider the symmetries of these equations.
The solutions are unchanged by the independent scalings
$\epsilon\to\lambda\epsilon$ and $\tau\to\zeta\tau$. Introduce the
variables $e=\ln(\epsilon/\epsilon_0)$ and $\theta=\ln(\tau/\tau_0)$,
where the arbitrary scales $\epsilon_0$ and $\tau_0$ can be chosen
to be the initial conditions. This removes the freedom of
scaling, so that the solutions of the equations can be written in the
form $e(\theta,\eta)$ and $y(\theta,\eta)$.  Thus, the scale symmetries
of the ideal hydrodynamic equations connect solutions with different
initial conditions.

Non-ideal hydrodynamics breaks both these symmetries by the
introduction of the time scale $\tau_\pi$ and the scale of
energy density $\varpi=\vis/\tau_\pi$. In other words, the scaling
$\epsilon\to\lambda\epsilon$  (simultaneously $\pi_V\to\lambda\pi_V$)
and $\tau\to\zeta\tau$ are not symmetries unless one simultaneously
scales $\tau_\pi\to\zeta\tau_\pi$ and $\vis\to\lambda\zeta\vis$. Thus,
the scalings relate flows of fluids with different material properties.
This is the relativistic analogue of scaling laws called ``physical
similarity'' \cite{landau} that one finds in non-relativistic
fluids. Such similarities are the basis of scaling invariants, also
known as dimensionless variables, such as the Reynolds number, $\rey$,
which are used to relate flows of different fluids.

The analysis here gives three scaling invariants---
\beq
   \chi = \frac\epsilon\varpi,\qquad
   \varphi = \frac{\pi_V}\varpi,\qquad{\rm and}\qquad
   \S = \frac{\pi_V}\epsilon.
\label{invs}\eeq
The dimensionless ratios lead to physical similarities between flows. We
can relate these variables with quantities familiar from Navier-Stokes
hydrodynamics by examining what they become in the appropriate limit.

As pointed out earlier, the ELNS limit of eqs.\ (\ref{hydro}) is obtained
when $\tau_\pi=0$. In that case, $\pi_V=4\vis\Theta/3$.
The quantity $\Theta$ is the inverse of a characteristic scale for the
flow, $\tau_c$. In the non-relativistic, \ie, the Navier-Stokes, limit,
$\tau_c$ is a characteristic time scale. In this limit we can define a
characteristic length scale for the flow by the relation $L_c=\tau_c v$,
where $v$ is the flow velocity. Then, one finds
\beq
   \S=\frac{4\vis}{3\epsilon\tau_c}=\frac{4c_s^2}3\left(\frac v{c_s}\right)^2
     \frac{\vis}{\epsilon L_cv} = \frac{4c_s^2}3 \frac{M^2}{\rey},
\label{oneid}\eeq
where $M=v/c_s$ is the Mach number of the flow and $\rey=\epsilon L_c
v/\vis$ is the Reynolds number. The first of the expressions on the
right comes from taking the ELNS limit, whereas the last expression
involves taking, additionally, the non-relativistic limit. In
Navier-Stokes hydrodynamics the ratio of inertial and viscous forces
is $\rey$ \cite{landau}. In this case one may therefore suspect that
$\S\propto1/\rey$. The exact relation above bears this out, with
corrections needed to translate between the fully relativistic and
non-relativistic formulas.  Similarly, one finds that
\beq
   \varphi = \frac{4\tau_\pi}{3\tau_c} = \frac4{3\xi}\,
       \frac{\lambda}{L_c}\, \frac{v}{c_s} = \frac4{3\xi}\,MK
\label{twoid}\eeq
where a mean-free path, $\lambda=\xi c_s\tau_\pi$, $\xi$ is
some numerical constant, and $K=\lambda/L_c$ is the Knudsen number. 

The third variable
\beq
   \chi=\frac\varphi\S=\frac1{c_s^2\xi}\,\frac{K\rey}M
\label{threeid}\eeq
is interesting, since the combination $K\rey/M\simeq1$ in Navier-Stokes
theory. For Boltzmann and conformal fluids this combination is
constant. The Navier-Stokes relation is obtained for these fluids when
$\xi$ is chosen appropriately. For the Boltzmann fluid, this happens when
$\xi\simeq2/3$.  Eqs.\ (\ref{oneid}, \ref{twoid}, \ref{threeid}) provide the
connection between $\S$, $\varphi$ and $\chi$ and $\rey$, $M$ and $K$ 
in the appropriate limit.

In ideal hydrodynamics thermalization and freezeout are notions which
are imposed from the outside. In non-ideal hydrodynamics, however, some
understanding of these phenomena could be possible \cite{outofeq}. In
the Navier-Stokes theory, for example, $K\simeq M/\rey$ must be smaller
than unity in order for the solutions to describe valid flows. Since a
solution of the Navier-Stokes equation allows us to compute both $M$
and $\rey$, one can use the solution to compute $K$ and determine
its own validity. A solution of the equations of IS hydrodynamics
gives $\tau_c=1/\Theta$. When $\tau_c$ is larger
than $\tau_\pi$, \ie, when $\varphi<4/3$, the solution corresponds to a
physical flow. In the case of scaling flow (discussed below), this gives
an initial time, $\tau_0$, at which the solutions begin to describe
physical fluid flows. Thus we have a self-consistent description of
thermalization. For scaling flow, $\varphi$ decreases with time. Hence,
after thermalization, IS hydrodynamics is always applicable; \ie,
we lack a description of freezeout. That phenomenon requires us to
examine radial flow. As argued before, radial flow becomes important
at a time $\tau_T=R/c_s$, where $R$ is the transverse size. Thus,
one-dimensional IS hydrodynamics in the scaling approximation is expected
to be valid in the range $\tau_0\le\tau\le\tau_T$. A description of
freezeout at late times has to be sought in the full 3-dimensional
hydrodynamics.

\subsection{Scaling solutions and fluctuations}

Solutions with $y=0$ are called scaling solutions or boost-invariant
solutions. It was argued by Bjorken that asymptotic freedom implies
that, at sufficiently high energies, hadron multiplicities must become
invariant under longitudinal boosts. Assuming further that these
multiplicities have their origin in the hydrodynamic distribution of
the entropy density, he argued that the relevant flows in high energy
heavy-ion collisions must be boost invariant \cite{bjorken}. While the
phenomenological relevance of this argument may be questioned, boost
invariance is simply analyzed.  One uses $y=0$, as a consequence of
which $S=0$ and $\Theta=1/\tau$. Substituting these into eqs.\
(\ref{hydro}), one finds that
\beq
    \tau D\epsilon = \pi_V-B\epsilon,\qquad
    \D\pi_V = c_s^2\D\epsilon,\qquad
    \tau D\pi_V = -\frac\tau{\tau_\pi} \pi_V + \frac43 \varpi,
\label{scaling}\eeq
where $D=\partial_\tau$ and $\D=(1/\tau)\partial_\eta$ (see
eq.\ \ref{derivonscalar}). Now,
Fourier transforming in $\eta$ decouples the Fourier
modes, labeled by $k$. The second equation implies that the
identity $\pi_V(\tau,k)=c_s^2\epsilon(\tau,k)$ must hold for all
$k\ne0$. However the other two equations cannot be manipulated to give
$D(\pi_V/\epsilon)=0$. Consequently, only the $k=0$ mode is allowed
to be non-zero, \ie, the solutions to these three equations must have
both $\epsilon$ and $\pi_V$ independent of $\eta$. This demonstrates
the well-known equivalence of the conditions of boost invariance and $y=0$.
Of course, the second of the three equations above becomes redundant and
the problem can be treated with the remaining two equations.

In the next few sections we will investigate the scaling solutions for
fluids with various different constitutive equations, and analyze their
stability \cite{baym,kouno}. Some numerical studies of the correlations
of fluctuations were reported in \cite{romatschke}. We note that the IS
hydrodynamic equations can be written in
the form $\mathbf x_\theta = \mathbf f(\theta,\mathbf x,\mathbf x_\eta)$,
where $\mathbf x$ is the vector of three hydrodynamic variables and
$\theta$ is the ``time'' variable. Now,
setting to zero the component corresponding to $y$ in $\mathbf x$
results in the boost-invariant equations, whose solutions we represent
by $\mathbf x^0$. Represent the fluctuations around this solution by
$\Delta\mathbf x^1$, where $\Delta\ll1$ is a dimensionless book-keeping
parameter. The equations for these fluctuations can be written to linear
order in $\Delta$,
\beq
   \mathbf x^1_\theta = M_0(\theta,\mathbf x^0)\mathbf x^1 +
        M_1(\theta,\mathbf x^0)\mathbf x^1_\eta,
\label{fluct}\eeq
where $M_0$ and $M_1$ are the Jacobian matrices of the derivatives of
$\mathbf f$ with respect to $\mathbf x$ and $\mathbf x_\eta$ respectively.
For the stability analysis one asks whether a given $\mathbf x^1$
increases or decreases with time.  Fourier transforming in $\eta$
decouples the derivatives with respect to the variables and gives
independent linear evolution equations for each mode--- $\mathbf
x^1_\theta(\theta,k) = M\mathbf x^1(\theta,k)$, where $M=M_0+ikM_1$. The
question of stability then reduces to examining $M$
and checking whether the solutions for $\mathbf x^1$ decrease faster
than the scaling solution or not.  We demonstrate the method with the
ideal fluid in the next section.

\section{Ideal fluid: the Bjorken solution and sound waves}

 \begin{figure}[htb]\begin{center}
    \scalebox{0.707}{\includegraphics{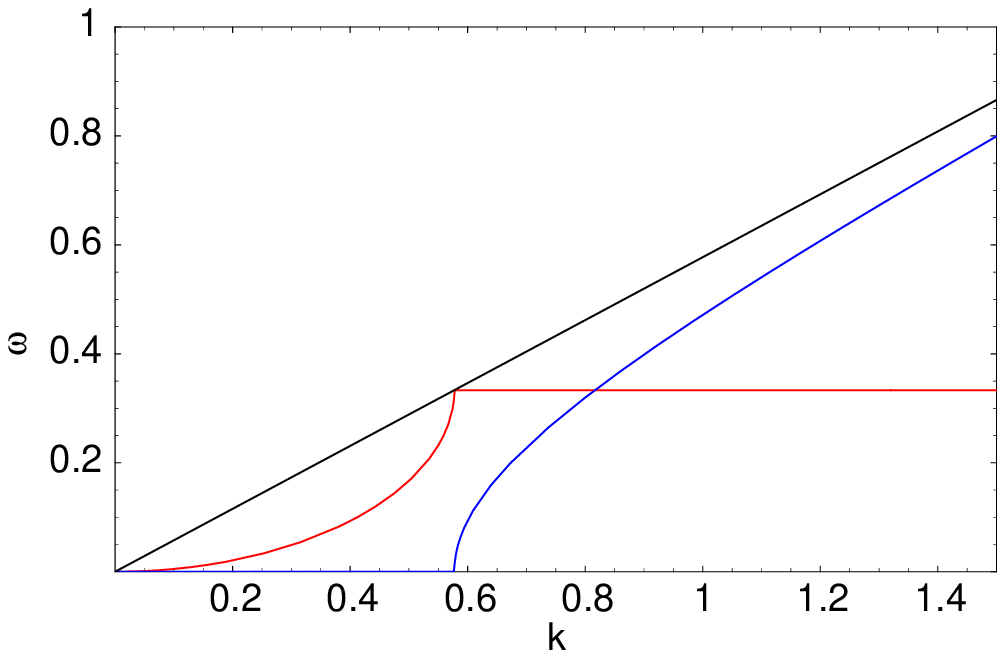}}
    \scalebox{0.707}{\includegraphics{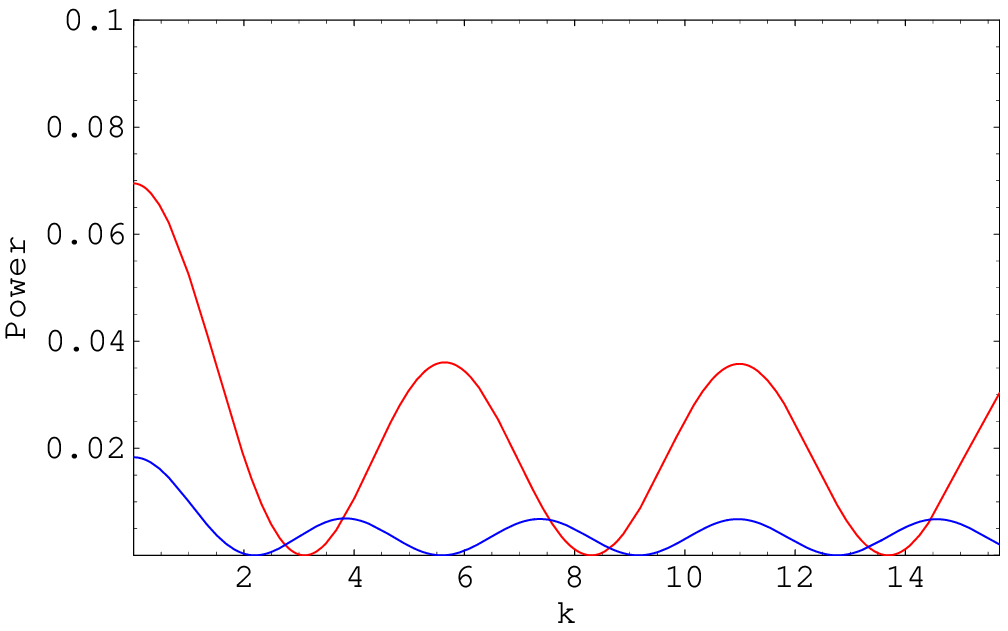}}
    \end{center}
   \caption{(Color online) On the left is the dispersion relation for sound waves
   in the Bjorken solution for $c_s^2=1/3$. The real part of $\omega$
   is shown in blue as a function of $k$ and the damping exponent, \ie,
   the imaginary part, is in red (the straight line in black is the
   line $\omega=c_s k$). There are no propagating waves for
   $k<1/\sqrt3$. On the right is the power spectrum resulting from the
   evolution of $P_\epsilon(\theta=0;k)=1$ at $\theta=1$ (upper
   curve) and 3/2 (lower curve).}
 \label{fg.idealdispersion}\end{figure}

The scaling solution in ideal hydrodynamics is obtained by setting
$\vis=\pi_V=0$ in eqs.\ (\ref{scaling}).  Introducing the variables
$e=\ln(\epsilon/\epsilon_0)$ and $\theta=\ln(\tau/\tau_0)$,  the
equations become---
\beq
   \frac{de}{d\eta} = 0, \qquad{\rm and}\qquad
    \frac{de}{d\theta} = - B.
\label{scalingeqs}\eeq
This gives the Bjorken solution
\beq
   e(\theta,\eta) = - B\theta, \qquad i.e.,\qquad
   \epsilon(\tau,\eta) = \epsilon_0 \left(\frac{\tau_0}\tau\right)^B.
\label{bjorken}\eeq
The entropy density is defined as $s=(\epsilon+p)/T$. Entropy is conserved
in an ideal fluid \cite{landau}. All this is well known.

The linearized equations for sound waves are extracted
by introducing a dimensionless parameter $\Delta\ll1$, which sets the scale
of quantities involved in the propagation of sound relative to the 
boost-invariant background, as discussed earlier. Then one proceeds by setting 
$y(\theta,\eta)=\Delta y^1(\theta,\eta)$ and writing
$e(\theta,\eta)=e^0(\theta)+\Delta e^1(\theta,\eta)$. Inserting these
into the ideal fluid equations (\ref{ideal}), one finds to order $\Delta^0$,
the equations (\ref{scalingeqs}). At order $\Delta$ one finds---
\beq
   e^1_\theta = - By^1_\eta,\quad{\rm and}\quad
   y^1_\theta = - \left(1-\frac1B\right) e^1_\eta + (B-2) y^1.
\label{idealsound}\eeq
A Fourier transformation in $\eta$ reduces the partial differential
equations above to uncoupled evolution equations for each Fourier mode.
Since these linear first-order evolution equations are autonomous,
\ie, they do not involve the variable $\theta$ anywhere explicitly,
the solutions are in the form of waves $\exp[i(\omega\theta+k\eta)]$.

Before proceeding, a point about wave solutions may need comment. 
Plane waves in Galilean coordinates have the form
$\exp[i(\omega t+\mathbf k\cdot\mathbf x)]$, unlike the one above.
However, the functional form of a wave solution is dependent on
the geometry of the situation; for example,
in spherical geometry a wave solution has the form $\exp[i(\omega
t+k|\mathbf x|)]/|\mathbf x|$. 
Wave solutions in boost-invariant geometry have the specific form
\beq
   \mathrm e^{i(\omega\theta+k\eta)} = 
         \left(\frac\tau{\tau_0}\right)^{i\omega}\mathrm e^{ik\eta} =
      \left(\frac{t+z}{\tau_0}\right)^{i(\omega+k)/2}
      \left(\frac{t-z}{\tau_0}\right)^{i(\omega-k)/2}.
\label{boostedwave}\eeq
A real value for $\omega$ corresponds to an oscillatory
solution.  In the form on the right it is manifest that the real parts of
$\omega$ and $k$ are not inverses of typical time and length scales for
oscillation. If $\theta_n$ and $\theta_{n-1}$ are, respectively, the $n$-th
and $n-1$st times that the phase of the wave repeats at a given spacetime
rapidity, then clearly $\theta_n-\theta_{n-1}=2\pi/\omega$, so that
$\tau_n=\tau_{n-1}\exp(2\pi/\omega)$.  In other words, the ``period'' of
oscillation increases geometrically in the number of periods. The $n$-th
time that the phase recurs after the initial time $\tau_0$ is given by
$\tau_n=\tau_0\exp(2\pi n/\omega)$. In the same way, at fixed $t$, the
``wavelength'' increases with $z$. Both
these scalings are direct consequences of boost-invariant expansion---
the longitudinal expansion red shifts sound waves. The analogy with the
red shifting of light in an expanding universe is clear; both follow
from the fact that the spatial components of the metric depend on the
time. If $\omega$ is complex, then the real part gives oscillations exactly
as described above. The imaginary part of $\omega$, \ie, $-\re i\omega$,
gives rise to power law behaviour in $\tau$. The scaling solution is
stable when $\re i\omega<0$, so that fluctuations are damped. Hence we
will give the name damping exponent to $-\re i\omega$.

Substituting the form of the wave solution in eq.\ (\ref{boostedwave}) 
into eq.\ (\ref{idealsound}), one obtains the dispersion relations
\beq
   i\omega = -\frac12 (1-c_s^2) \pm \frac12\sqrt{4c_s^2(k_0^2-k^2)}
     \qquad{\rm where}\qquad
   k_0 = \frac{1-c_s^2}{2c_s}.
\label{dispideal}\eeq
For $k<k_0$ the modes are overdamped; the two damping exponents
are equally spaced around $(1-c_s^2)/2$. Only for $k>k_0$, are there
propagating modes. These are damped due to the expansion of the scaling
solution. 
The slowest decreasing part corresponds to the positive sign above. As
a result, this is the part that is visible to experiments.
This dispersion relation is shown in Figure \ref{fg.idealdispersion}.
A general solution for $e^1$ can be written in the form
\beq
   e^1(\theta,k)=c_+ \mathrm e^{i\omega_+\theta} +
    c_- \mathrm e^{i\omega_-\theta},
\label{gensolidac}\eeq
where $c_\pm$ depend on the initial conditions, and $\omega_\pm$
are the solutions in eq.\ (\ref{dispideal}) with the corresponding signs.
Similar solutions can be written for $y^1$. Since the real parts
of $i\omega$ are non-positive, the fluctuations do not grow, and the
scaling solution is stable \cite{baym,kouno}.

The physics of sound can be captured in the evolution of the power
spectrum of fluctuations of the energy density,
\beq
   P_\epsilon(\tau;k) = \left|\epsilon^1(\tau;k)\right|^2
   \quad{\rm where}\quad
   \epsilon^1(\tau;k) = \int\frac{d\eta}{\sqrt{2\pi}}
     \mathrm e^{-ik\eta} \epsilon^1(\tau,\eta).
\label{powerspec}\eeq
Since $\epsilon^1(\tau,\eta)=e^1(\tau,\eta)\epsilon^0(\tau)$, for the
ideal fluid one may write asymptotically, when the component in $c_-$
can be neglected,
\beq
   P_\epsilon(\tau;k) = P_\epsilon(\tau_0;k)
      \left(\frac{\tau_0}\tau\right)^{2\omega_d},
    \quad{\rm where}\quad
   \omega_d(k)=B+\frac12(1-c_s^2)-c_s\sqrt{k_0^2-k^2}{\mathbf H}(k_0-k),
\label{idealpower}\eeq
$k_0$ is given in eq.\ (\ref{dispideal}), and $\mathbf H$ denotes the
unit step function. At not so late times, the interference between the
frequencies $\omega_+$ and $\omega_-$ (when $k>k_0$) gives rise to beats.
The shape of the power
spectrum resulting from an initially flat power spectrum of fluctuations
through exact solution of eqs.\ (\ref{idealsound}) with initial conditions
$c_+=c_-$ is shown in Figure \ref{fg.idealdispersion}.  At short times
it is dominated by beats.  The expression in eq.\ (\ref{idealpower})
is asymptotic. As shown in Figure \ref{fg.idealdispersion}, it is not
recovered for $\theta=3/2$.  It is clear that if the initial conditions
contain fluctuations around the boost-invariant values, then after
sufficient evolution, these fluctuations are damped.  The longer the
duration of hydrodynamic evolution, the more damped the fluctuations.
A detailed analysis of the growth of transients at short times is given
in Appendix C.

If the initial conditions, \ie, the power spectrum at time $\tau_0$,
for an ideal fluid were known, then an event-by-event measurement of
the power spectrum of the acoustic energy density would be able
extract the value of $c_s$ and thereby give a measurement of the
equation of state. Conversely, if the equation of state were known,
then the same data could be used to extract, event by event, the
initial conditions.

The power spectrum studied here is closely related to the correlation
function of fluctuations studied in \cite{romatschke}. In Fourier space
the correlation function corresponds to studying the joint distribution
of fluctuations at different $k$, whereas the power spectrum gives the
variance in the fluctuations at a single $k$. Sonic peaks are visible
in both the quantities. The analysis of the power spectrum in terms of
interference, and the consequent clear relation with $c_s$, is a little
harder to establish for the correlation function.

\section{A simple fluid}

The simple non-ideal fluid model defined in Section II.A has constant values
of all constitutive parameters--- $c_s$, $\tau_\pi$ and $\vis$. Using
the variables $\chi$ and $\varphi$ introduced in eq.\ (\ref{invs}) and
$\vartheta= \tau/\tau_\pi$, we make the decomposition---
\beq
   \chi(\vartheta,\eta) = \chi^0(\vartheta) + \Delta \chi^1(\vartheta,\eta),
    \qquad
   \varphi(\vartheta,\eta) = \varphi^0(\vartheta) + \Delta \varphi^1(\vartheta,\eta),
    \qquad
   y(\vartheta,\eta) = \Delta y^1(\vartheta,\eta).
\label{decom}\eeq
These expansions have to be substituted into the eqs.\ (\ref{hydro}) and
the material properties of the simple fluid used to extract equations for
the boost-invariant solution $\chi^0$ and $\varphi^0$, and the fluctuations
$\chi^1$, $y^1$ and $\varphi^1$.
In the next two subsections we examine
these two problems.

\subsection{The scaling solution}

The equations satisfied by the scaling solution are---
\beq
   \frac{d\chi^0}{d\vartheta} =
       -\frac B\vartheta\chi^0 + \frac{\varphi^0}\vartheta
     \qquad{\rm and}\qquad
   \frac{d\varphi^0}{d\vartheta} = -\varphi^0 + \frac4{3\vartheta}.
\label{simpleis}\eeq
One solves the second equation and then inserts the solution into
the first. It is easy to check that
\beq
   \varphi^0(\vartheta) = {\rm e}^{-\vartheta}
      \left[b +\frac43 {\rm Ei}(\vartheta)\right],\qquad
   b = -\frac43{\rm Ei}(\vartheta_0)+\mathrm e^{\vartheta_0}
            \varphi^0(\vartheta_0),
\label{solveone}\eeq
where ${\rm Ei}(x)$ denotes the exponential integral \cite{grad},
and $\vartheta_0=\tau_0/\tau_\pi$. We will use $\vartheta_0=1$ in
numerical work. The asymptotic expansion of the exponential integral,
\beq
   {\rm Ei}(x) \sim \frac{{\rm e}^x}x\left(1+\frac1x\right),
\eeq
can be used to write down the asymptotic expression---
\beq
   \varphi^0(\vartheta) \sim b {\rm e}^{-\vartheta} +
        \frac4{3\vartheta}\left(1+\frac1\vartheta\right).
\label{solveoneasy}\eeq
Note that the solution of the homogeneous equation decays much
faster than the particular integral. Hence, the long-time behaviour
of $\pi_V$ is nearly independent of the initial conditions on this
quantity.

 \begin{figure}[htb]\begin{center}
    \scalebox{0.5}{\includegraphics{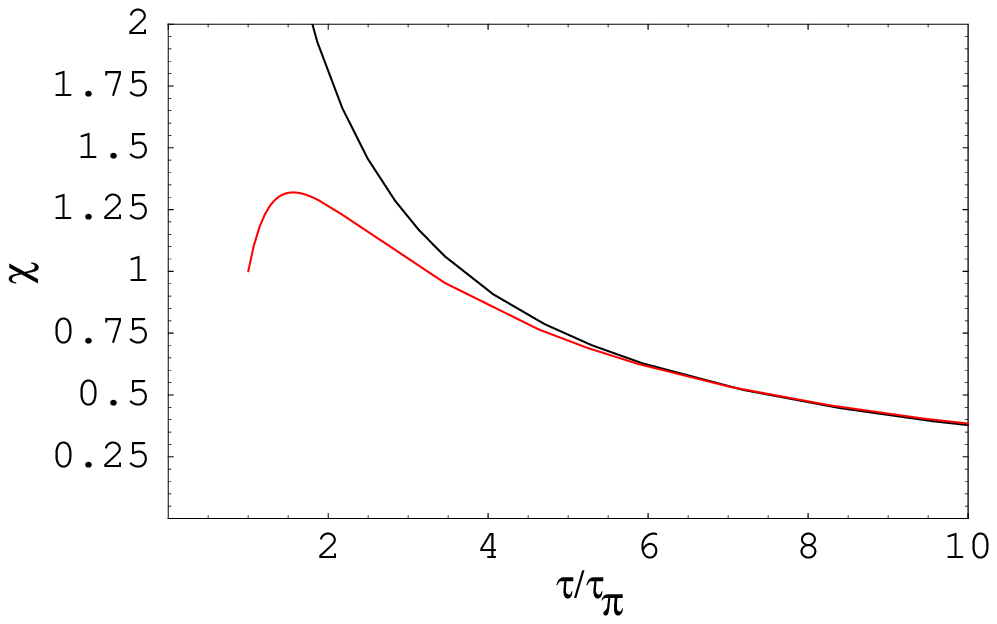}}
    \scalebox{0.5}{\includegraphics{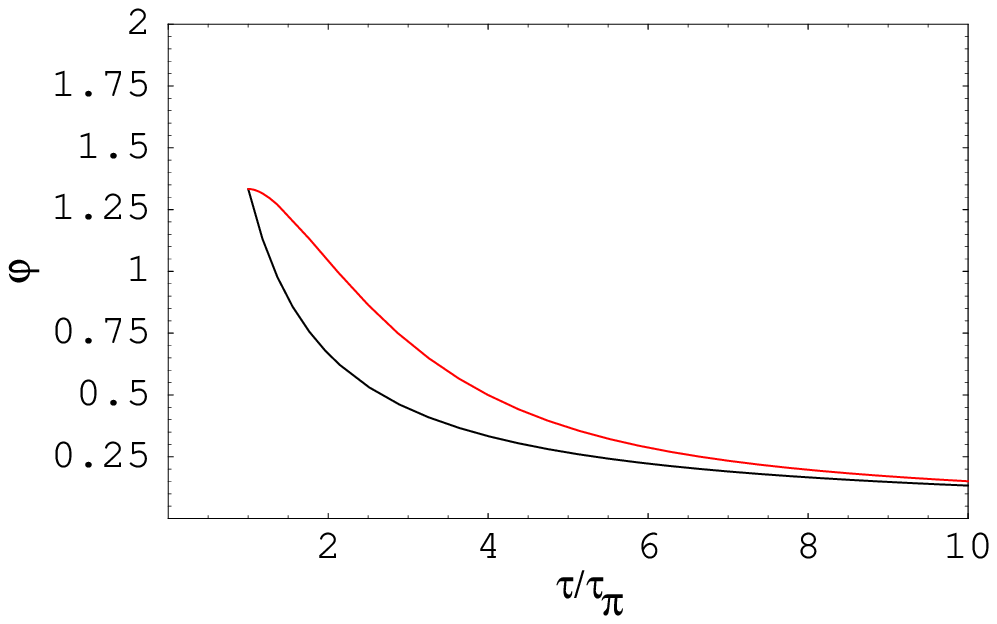}}
    \scalebox{0.5}{\includegraphics{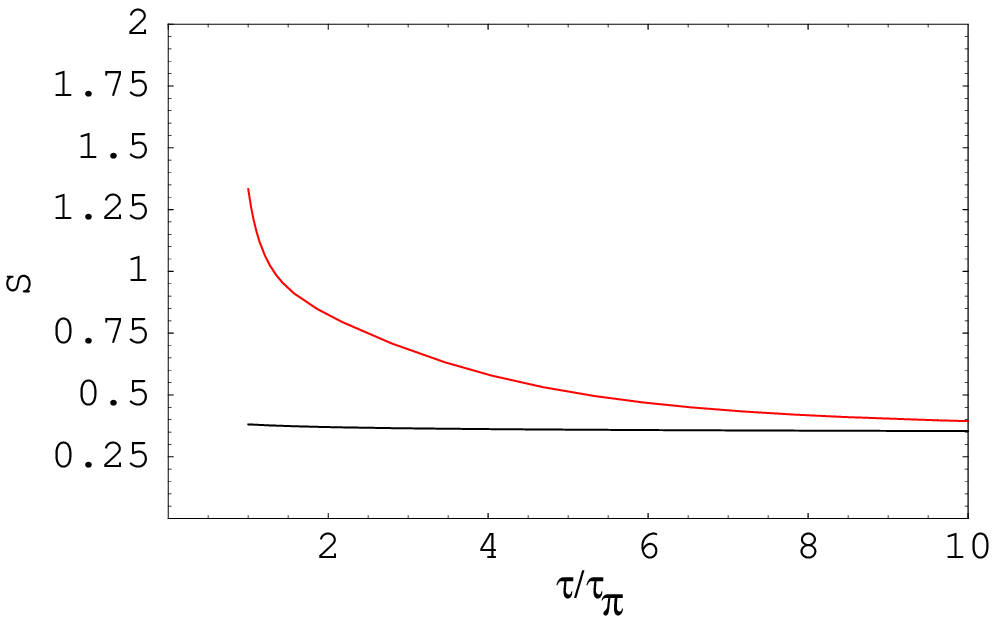}}
    \end{center}
   \caption{(Color online) Comparison of ELNS hydrodynamics (lines in black) and 
   IS hydrodynamics (in red). The evolution of normalized energy
   density, $\chi^0$, normalized shear stress, $\varphi^0$, and
   $\S=\varphi^0/\chi^0$, are shown as a function of the normalized
   time, $\tau/\tau_\pi$. The initial condition $\varphi^0(1)=4/3$
   in both cases.}
 \label{fg.elnsvsis}\end{figure}

Inserting the solution in eq.\ (\ref{solveone}) into the first of eq.\
(\ref{simpleis}) gives a solution in terms of the Meijer G-functions
\cite{grad,wolfram}.  Instead, by inserting the asymptotic expansion
in eq.\ (\ref{solveoneasy}) into the equation, we can find a simplified
solution---
\beq
   \chi^0(\vartheta) = \frac a{\vartheta^B} -
      b \frac{\Gamma(B,\vartheta)}{\vartheta^B} +
         \frac4{3c_s^2 \vartheta} + \cdots
\label{final}\eeq
where $a$ is determined by the initial condition $\chi^0(\vartheta_0)$
and the remaining terms decay as higher integer powers of $1/\vartheta$.
The incomplete Gamma function is defined to have the values $\Gamma(B,0) =
\Gamma(B)$ and $\Gamma(B,\infty) = 0$.  The solution of the homogeneous
equation gives the Bjorken solution.  However, this is not the slowest
falling part; that is given by the $1/\vartheta$ term, which arises
from the inhomogeneous part of the equation, and has no dependence
on initial conditions.  In the very late time limit, one finds
$\varphi^0\simeq4/3\vartheta$ and $\chi^0\simeq4/3c_s^2\vartheta$, so that
the Bjorken solution is never recovered. The expansion is slowed down with
respect to the ideal fluid because of dissipation--- in expanding against
the vacuum, an ideal fluid pumps all its energy into the expansion,
whereas a non-ideal fluid dissipates energy, thus slowing the expansion.
Similar results were also obtained in \cite{nonidealhic} for the case of
ELNS hydrodynamics.

\subsubsection{Comparison of ELNS and IS hydrodynamics}

The differences between ELNS and IS hydrodynamics are illuminating. The
ELNS limit of IS hydrodynamics is obtained formally by setting
$\tau_\pi=0$, or, equivalently, by dropping the term $D\pi_V$ in eqs.\
(\ref{hydro}). Now, a comparison of the two theories makes sense only
after appropriate normalization of both time and energy. So in ELNS
hydrodynamics one must introduce artificially a unit of time which is
numerically equal to the value of $\tau_\pi$ used in IS hydrodynamics.

Then, rewriting the ELNS equations, one obtains scaled quantities
$\chi^0$ and $\varphi^0$ which are directly comparable to the same
quantities in IS. The equations of ELNS are then obtained from eqs.\
(\ref{simpleis}) by dropping the term $d\varphi^0/d\vartheta$, so that
$\varphi^0=4/3\vartheta$. Inserting this into the equation for $\chi^0$,
one has 
\beq
   \frac{d\chi^0}{d\vartheta} = -\frac{B\chi^0}\vartheta +
         \frac4{3\vartheta^2},
    \qquad{\rm so}\qquad
   \chi^0(\vartheta) = \frac4{3(B-1)\vartheta} + 
          \frac{\chi^0(1)-4/3(B-1)}{\vartheta^B}.
\label{elnssimplechi}\eeq
A comparison of the resulting solutions is shown in Figure
\ref{fg.elnsvsis}.  The initial conditions for $\varphi^0$ are chosen to
be the same; in IS hydrodynamics it corresponds to the smallest time at
which the equations are valid, \ie, $\tau_0=\tau_\pi$. 
As in \cite{muronga,baier}, we find clear differences
between ELNS and IS evolution. In earlier works the difference in the
evolution of the energy density was investigated for equal initial conditions.
We have chosen the
initial conditions for $\chi^0$ to be different, but tuned so that the
ELNS and IS solutions approach each other at large times. Since initial
conditions are not detectable in heavy-ion collisions, it is of interest
to see that large-time behaviour cannot, by itself, distinguish between
ELNS and IS hydrodynamics. Moreover, the same final energy density can
lead to different estimates of the initial energy density in the two
kinds of dynamics.

\subsubsection{Entropy production}

Assuming that $s=\gamma\epsilon^{3/4}$, where $\gamma$ is a dimensionless 
quantity which is almost temperature independent,
$\sigma=(\chi^0)^{3/4}$ is a dimensionless quantity proportional to the
entropy density. The first of eqs.\ (\ref{simpleis}) can be
easily manipulated into the form---
\beq
   \frac{d\sigma}{d\vartheta} = -\frac\sigma\vartheta
              +\frac{3\varphi^0}{4\vartheta\sigma^{1/3}}.
\label{entropydensity}\eeq
Using the solutions for $\varphi^0$ and $\chi^0$ in eqs.\ (\ref{solveoneasy},
\ref{final}), one finds that at late times the right hand side is
negative.  This is in accord with the exact solution for $\chi^0$
shown in Figure \ref{fg.elnsvsis}. Depending on the initial conditions,
$\sigma$ may increase initially. However, at sufficiently large time it must
decrease. If there is initial growth in $\sigma$, then the turnover comes
when the right hand side of the above equation passes through zero, \ie,
at the value of $\vartheta$ when $\chi^0=3\varphi^0/4$.

Note, however, that the element of 3-volume contains a scale factor $\tau$
from the metric. This implies that the total entropy scales as $\Sigma
=\sigma\vartheta$. It is a straightforward exercise to rewrite eq.\
(\ref{entropydensity}) to obtain
\beq
   \frac{d\Sigma}{d\vartheta} = 
              \frac{3\varphi^0}4\left(\frac\vartheta\Sigma\right)^{1/3}.
\label{entropy}\eeq
The right hand side is manifestly positive definite, indicating that
the total entropy increases with time. Using the asymptotic expansion
of $\varphi^0$ in eq.\ (\ref{solveoneasy}), we find that $\Sigma$
grows asymptotically as $\vartheta^{1/4}$. One has the same power law
growth of $\Sigma$ in ELNS dynamics.

\subsection{Sound waves}

Using the decomposition of eq.\ (\ref{decom}), we examine
small fluctuations around the scaling solution.
At large $\vartheta$ we may use the asymptotic solutions
$\chi^0=4/3c_s^2\vartheta$ and $\varphi^0=4/3\vartheta$ in the
fluctuation equations. We also transform to the variable
$\theta=\ln\vartheta$ and use $y^1=g\exp\theta$. After Fourier
transforming in $\eta$, the equations for fluctuations
take the form---
\beqa
\nonumber
   &&\partial_\theta \left(\matrix{\chi^1\cr g\cr\varphi^1}\right) =
          M \left(\matrix{\chi^1\cr g\cr\varphi^1}\right),
      \qquad{\rm where}\qquad M = ikM_1+M_0,\\
     && \qquad\qquad M_1=\left(\matrix{
      0 & - \frac4{3c_s^2} & 0\cr
    -\frac{3c_s^4}4 & 0 & \frac{3c_s^2}4\cr
      0 & \frac43\mathrm e^\theta & 0}\right),\qquad
     M_0=\left(\matrix{ - B & 0 & 1\cr 0 & - 2 & 0\cr
      0 & 0 & - \mathrm e^\theta}\right).
\label{simplesetintheta}\eeqa
In contrast to the equations for fluctuations in an ideal fluid, \ie,
eqs.\ (\ref{idealsound}), these equations are not autonomous. As a
result, they cannot be solved by Fourier expansion in $\theta$. A
numerical solution is always possible, and we can examine the limits
of large and small $k$ analytically. At every $\tau$ there is an upper
cutoff on $k$ imposed by the requirement of the applicability of hydrodynamics.
This cutoff increases with $\tau$.

 \begin{figure}[htb]\begin{center}
    \scalebox{0.5}{\includegraphics{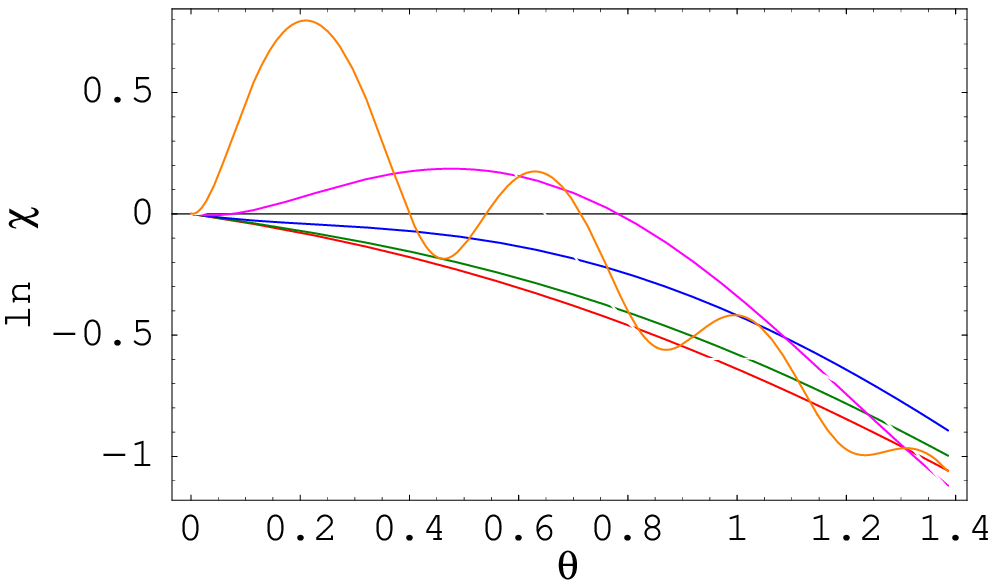}}
    \scalebox{0.5}{\includegraphics{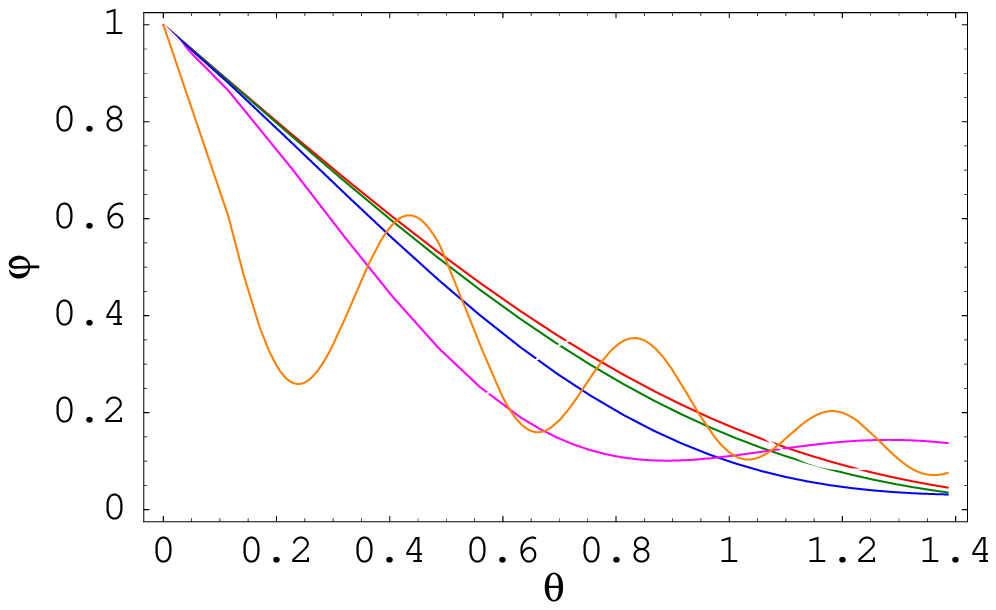}}
    \scalebox{0.5}{\includegraphics{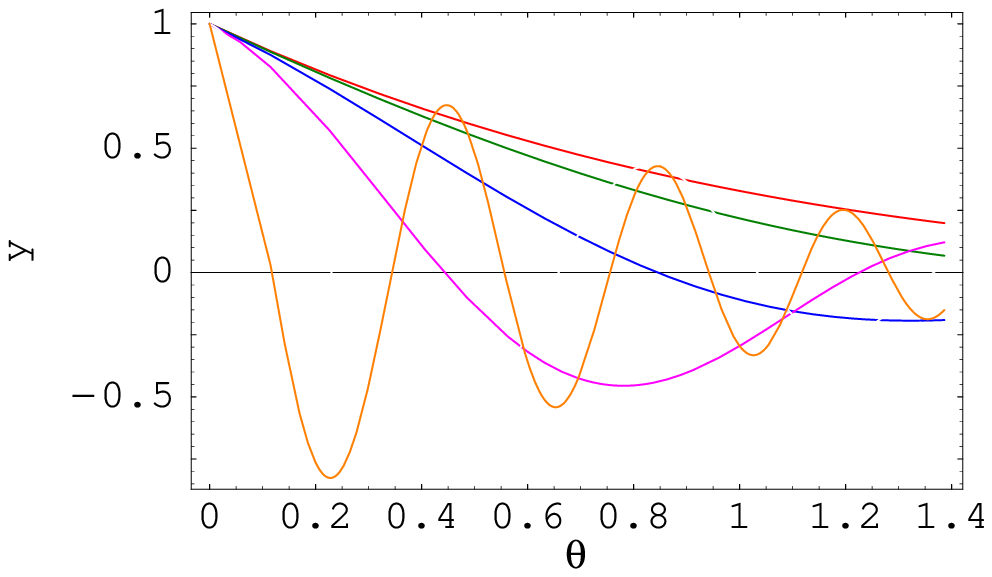}}
    \end{center}
   \caption{(Color online) Solutions of the linearized equations for fluctuations around
   the scaling solution in a simple fluid. The normalized energy, $\chi^1$,
   shear, $\varphi^1$, and $y^1$ are shown for $k=1/2$ (red), 1
   (green), 2 (blue), 4 (purple) and 8 (orange). The first three values
   of $k$ are overdamped, but the last two show oscillatory behaviour. The
   frequency of oscillation increases with $k$.}
 \label{fg.simplesound}\end{figure}

 \begin{figure}[htb]\begin{center}
    \scalebox{0.707}{\includegraphics{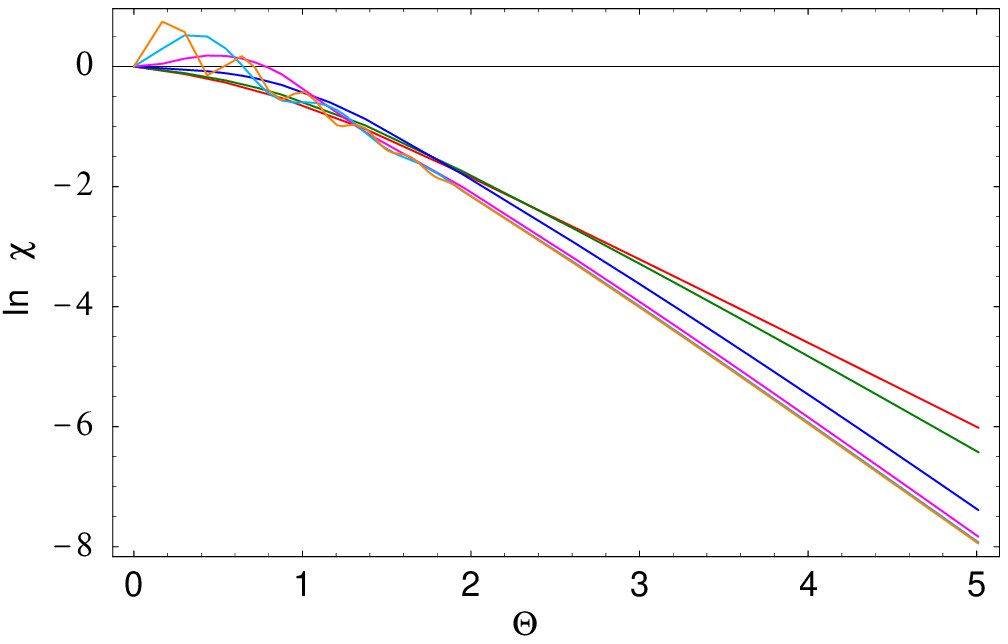}}
    \scalebox{0.707}{\includegraphics{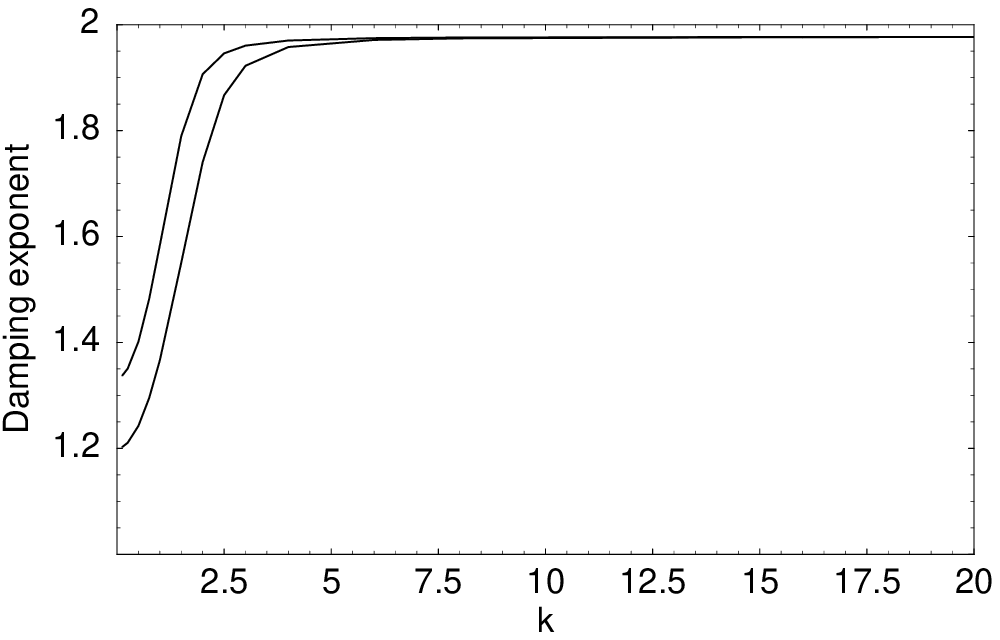}}
    \end{center}
   \caption{(Color online) The panel on the left shows $\ln\chi^1(k,\tau)$ as a function
   of $\theta=\ln(\tau/\tau_\pi)$ for the simple fluid with $c_s^2=1/3$
   for $k=1/2$ (red), 1 (green), 2 (blue), 4 (purple), 8 (light blue) and
   16 (orange). The panel on the right shows the damping exponent, \ie, the
   imaginary part of $\omega$, for $c_s^2=1/5$ and $1/3$, plotted as a
   function of $k$.}
 \label{fg.simpledamping}\end{figure}

In the limit $k\to0$, one may set $M=M_0$. Then, since $M_0$ is diagonal,
one can read off the solutions easily. $\chi^1$ and $g$ (and hence $y^1$)
are overdamped, whereas $\varphi^1$ decays exponentially in $\tau$. The
exact solutions are---
\beqa
\nonumber
   \chi^1(k\to0,\tau) &=& \chi^1(k\to0,\tau_0) \left(\frac{\tau_0}\tau\right)^B
      -\varphi^1(k\to0,\tau_0){\mathrm e}^{\tau_0/\tau_\pi}
               \left(\frac{\tau_\pi}\tau\right)^B
        \left[\Gamma\left(B,\frac\tau{\tau_\pi}\right)-
          \Gamma\left(B,\frac{\tau_0}{\tau_\pi}\right)\right],\\
\nonumber
   y^1(k\to0,\tau) &=& y^1(k\to0,\tau_0)\left(\frac{\tau_0}\tau\right),\\
   \varphi^1(k\to0,\tau) &=& \varphi^1(k\to0,\tau_0)
            \exp\left(\frac{\tau_0-\tau}{\tau_\pi}\right).
\label{sosmallk}\eeqa

In the regime $k\gg1$, a first approximation would be to neglect $M_0$.
As a result, one expects $\omega\propto k$, where $i\omega$ is an
eigenvalue of $M$.  Since $\omega$ is very large, $\exp\theta$ changes
little over many oscillations. Consequently one could treat this factor
as constant whenever it appears inside $M$. Within this approximation the
equations above can be treated as autonomous and therefore generically
describe oscillations.

 \begin{figure}[htb]\begin{center}
    \scalebox{1.0}{\includegraphics{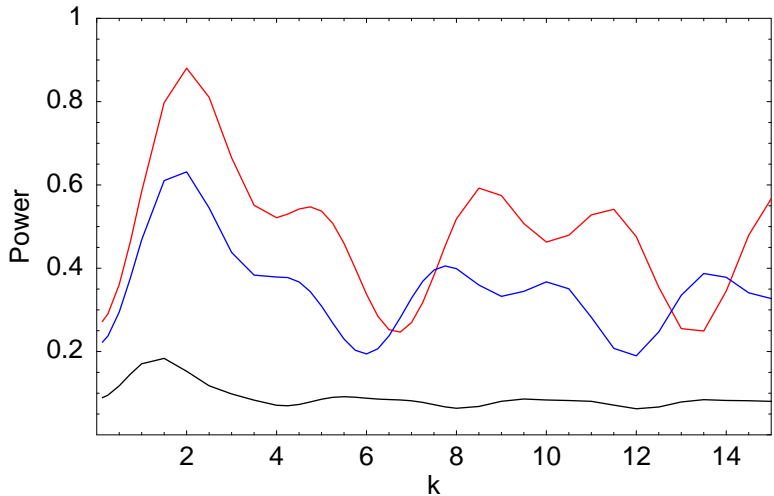}}
    \end{center}
   \caption{(Color online) The power spectrum obtained from an initially flat power
     spectrum of fluctuations at $\theta=1$ (red), 1.1 (blue) and
     1.5 (black). Here $c_s^2=1/3$ and $\tau_0$ has been set equal
     to $\tau_\pi$. Initial conditions are $\chi^1(0) = y^1(0)
     = \varphi^1(0) = 1$.}
 \label{fg.simplepower}\end{figure}

Since the eigenvalues of $M_1$ are real, within the approximation
where one neglects $M_0$, the solutions are not damped.  To go beyond
this and obtain the damping exponent, one sets up a perturbation theory
in powers of $1/k$ by writing $i\omega=ik\lambda_1+\lambda_0+{\cal
O}(1/k)$.  The simplest way to proceed is to substitute this form into
the characteristic equation for $M$. One finds then that this equation
has a leading term of order $k^3$ (which solves the eigenvalue problem
for $M_1$) and a first perturbation term of order $k^2$. The latter shows
growing solutions in the unphysical region $\exp\theta\ll1$, but damping
with $i\omega=-2+{\cal O}(1/k)$ when $\exp\theta\gg1$. Note that this
damping exponent is independent of $c_s^2$.

The numerical solutions to eqs.\ (\ref{simplesetintheta}) are shown for
a range of $k$ in Figure \ref{fg.simplesound}. There is clear evidence
of overdamped solutions for $k\le2$ and damped oscillatory solutions
for $k>2$.  In Figure \ref{fg.simpledamping} we show that $\chi^1$, which
is proportional to the energy density of fluctuations, is damped fairly
rapidly. However, for $\theta\le1$ there are clear signs of transients;
a detailed discussion of which is given in Appendix C.  Numerically we
see that $\chi^1$ is damped as a power of $\tau$ at large $\tau$,
making it easy to extract the damping exponent numerically. As shown
in the figure, and in agreement with our analysis above, at small $k$
this goes to $1+c_s^2$ and climbs to the neighbourhood of 2 at large $k$.

The power spectrum of fluctuations starting from an initially flat
spectrum, $P_\epsilon(0;k)=1$, is easily amenable to computation, and is
shown in Figure \ref{fg.simplepower}. The transient growth phenomenon
gives rise to several peaks for $\theta\le1$. By $\theta\simeq1.5$
the effect of the damping exponent is clearly visible. A complete
analysis is given in Appendix C.

\subsubsection{Sound in ELNS hydrodynamics}

 \begin{figure}[htb]\begin{center}
    \scalebox{1.0}{\includegraphics{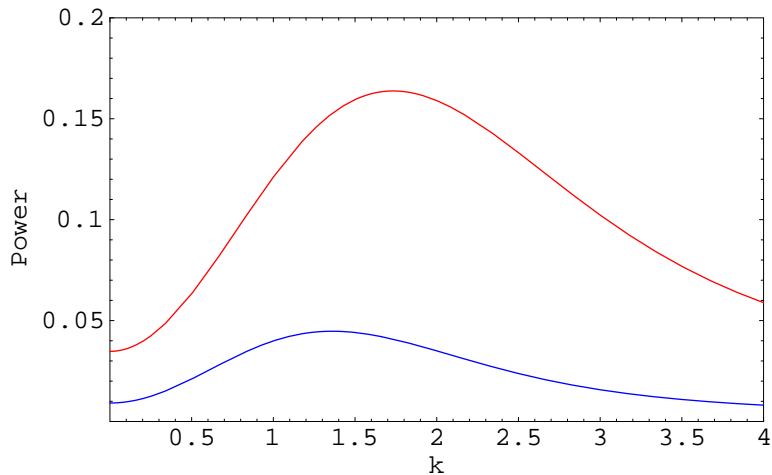}}
    \end{center}
   \caption{(Color online) The power spectrum obtained in ELNS hydrodynamics for a
     simple fluid from an initially flat power
     spectrum of fluctuations at $\theta=2$ (red) and
     2.5 (blue). Here $c_s^2=1/3$ and $\tau_0$ has been set equal
     to $\tau_\pi$.}
 \label{fg.elnssimplepower}\end{figure}

Sound propagation in ELNS hydrodynamics is quite different.  The
equations for sound can be obtained from eqs.\ (\ref{simplesetintheta}),
by simply dropping the term $d\varphi^1/d\theta$, \ie, by
setting $\varphi^1=4ikg/3$, and treating $\tau_\pi$
as an arbitrary scale of time, set equal to the value it would have
in IS hydrodynamics. When this relation is inserted into the equation
for $g_\theta$, a term in $g_{\eta\eta}$ is generated. This is
the diffusive term we expect in ELNS hydrodynamics, and arises directly
from the neglect of the relaxation time in shear pressure.

As before, we transform to variables $\theta=\ln\vartheta$ and write
$y^1=g\exp\theta$. The equations for sound in ELNS hydrodynamics are
\beq
   \partial_\theta \left(\matrix{\chi^1\cr g}\right) =
          M \left(\matrix{\chi^1\cr g}\right),
      \quad{\rm where}\quad M = -\left(\matrix{
          1+c_s^2 & \frac{4(1-c_s^2)ik}{3c_s^2}\cr
          \frac34ic_s^4k & 2+c_s^2k^2}\right).
\label{elnssimplesetintheta}\eeq
In the limit of small $k$, \ie, $k\ll1$, it is clear that
$\chi^1\propto\tau^{-B}$ and $y^1\propto1/\tau$. This is similar to the
results in eq.\ (\ref{sosmallk}). In this limit the solutions
of ELNS and IS hydrodynamics are not qualitatively different. This
is expected since the characteristic time scale of the wave is much
larger than $\tau_\pi$. 

In this case, an exact solution is possible for all $k$, since the equations
are autonomous. The eigenvalues of $M$ are
\beq
   i\omega=\cases{-(B+c_s^2 k^2),\cr -2,}
\label{elnssound}\eeq
where the combinations $4(1-c_s^2)\chi^1/3c_s^4-ikg$ and 
$4ik\chi^1/3c_s^2+g$ decay respectively with these damping
exponents. Not only is $y^1$ diffusively damped at large $k$,
but there are no propagating modes at all. This behaviour is
characteristic of parabolic equations.

In the absence of propagating modes there are no beats. Transient growth
can occur, but there is only a single peak in the power spectrum of the fluctuations
of energy.
The transient analysis is given in
Appendix C. The power spectrum resulting from an exact
numerical solution, starting from $P_\epsilon(0;k)=1$, is shown in Figure
\ref{fg.elnssimplepower}. Comparing this with Figure \ref{fg.simplepower},
we see that there is a clear difference between diffusive damping of
fluctuations in ELNS hydrodynamics and sound in IS dynamics.

\section{A Boltzmann fluid}

A Boltzmann fluid is defined by the constitutive relation $\chi =
\epsilon\tau_\pi/\vis=3/2c_s^2=9/2$, where, as discussed in Section
II.A, the three quantities $\epsilon$, $\vis$ and $\tau_\pi$, all
depend on the temperature. We rewrite the hydrodynamic equations
in terms of the variables $\S$, $y$, $u=T\tau/(T_0\tau_0)$, and
$\theta=\ln(\tau/\tau_0)$, where the initial conditions are
applied at the time $\tau_0$, \ie, at $\theta=0$.  We analyze
the scaling solution and its stability by the usual technique
of writing $u(\tau,\eta)=u^0(\tau)+\Delta u^1(\tau,\eta)$,
$\S(\tau,\eta)=\S^0(\tau)+\Delta\S^1(\tau,\eta)$ and $y(\tau,\eta)=\Delta
y^1(\tau,\eta)$. Substituting these into eqs.\ (\ref{hydro}), using the
material properties, and separating out the equations to orders $\Delta^0$
and $\Delta$, we obtain the equations which lead to the scaling solution
from the former, and the equations for fluctuations from the latter.

\subsection{The scaling solution}

The equations for the scaling flow become
\beq
   \partial_\theta\left(\matrix{u\cr\S}\right) = \V,\qquad{\rm where}\qquad
     \V=\left(\matrix{u\left[\S+3-c_s^2\right]/4 \cr
     \A - \S^2 + \S\left[(1+c_s^2) - \B u\right]}\right),
\label{bfluid}\eeq
$\A=4/3\chi$, $\B=\tau_0/\tau_\pi(0)$ and $c_s^2$ are non-negative.
In this subsection we lighten the notation by writing $u$ for $u^0$
and $\S$ for $\S^0$.
Three numbers are needed to fix the initial condition in the original
formulation of the problem (eq.\ \ref{hydro}), \ie, the initial time
$\tau_0$ and the values of $\epsilon(\tau_0)$ and $\pi_V(\tau_0)$.
The initial condition on $u$ is, by definition, $u(\theta=0)=1$. The
two free parameters in the initial conditions are transformed into the
value of $\S(0)$ and the value of the parameter $\B$ which appears in
the equation. Note that $\B<1$ is disfavoured. A dimensional quantity is
needed to complete the specification of the initial conditions, and we
can choose this to be $T_0=a\B/\tau_0$. Next, choosing $c_s^2=1/3$ (and
hence $\chi=9/2$), one has $\A=8/27$.  Since the equations are autonomous,
one can analyze them using the phase plane method \cite{arnold}.

\subsubsection{Phase plane structure}

 \begin{figure}[htb]\begin{center}
    \scalebox{0.32}{\includegraphics{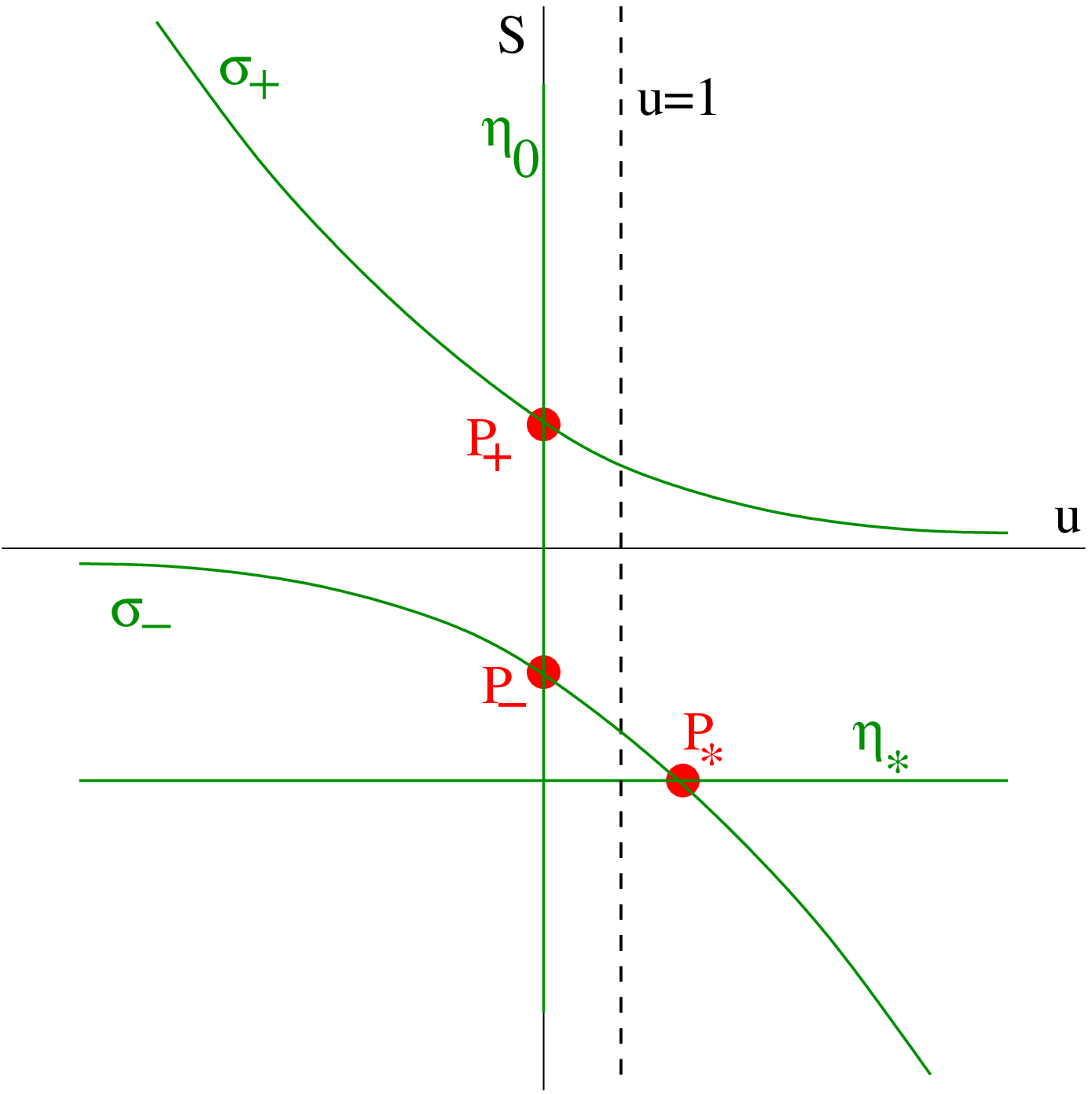}} \hfil
    \scalebox{0.5}{\includegraphics{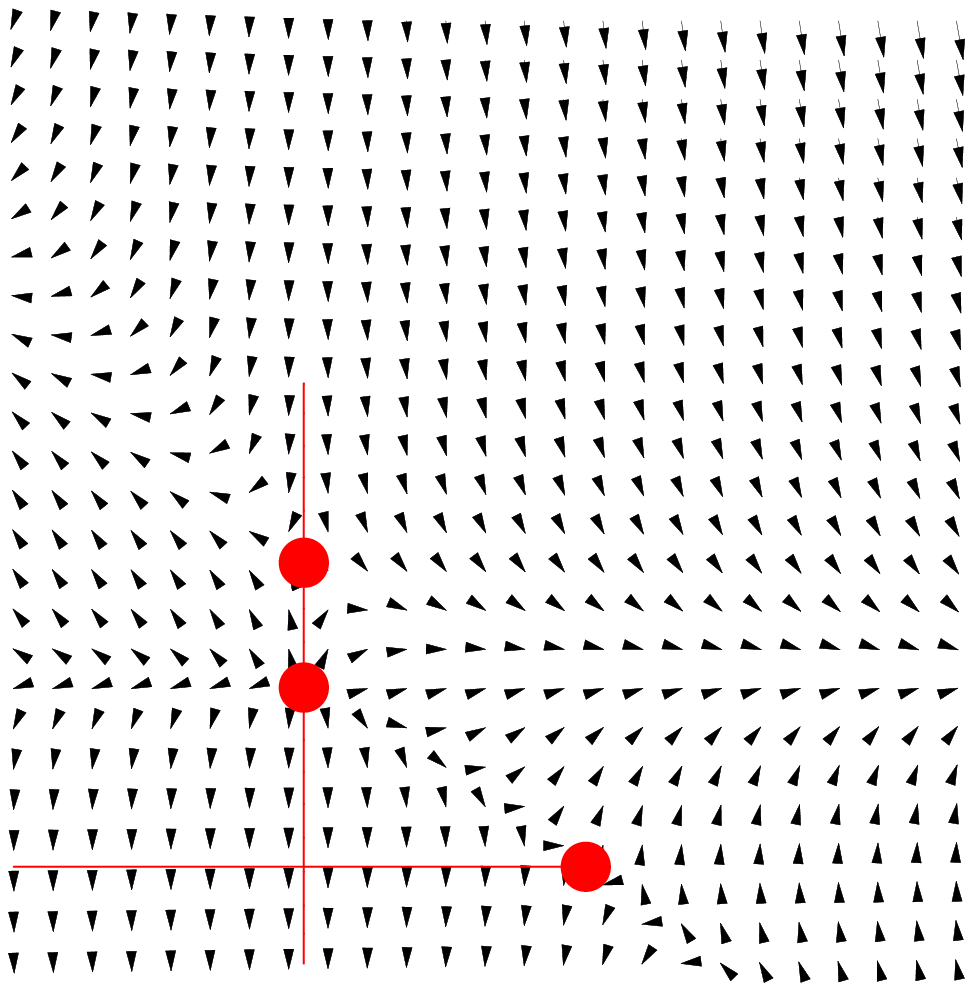}}
    \end{center}
   \caption{(Color online) The panel on the left shows structural elements of the
   phase plane: the nullclines, \ie, the lines along which one of
   the derivatives vanishes (that of $u$ vanishes on $\eta_0$ and
   $\eta_*$, that of $\S$ vanishes on $\sigma_\pm$), and the fixed
   points, at which both derivatives vanish. The panel on the right
   shows the vector field of derivatives and the fixed points.}
 \label{fg.flows}\end{figure}

The idea of the phase plane method is that the right
hand sides of eqs.\ (\ref{bfluid}) define the local direction of
derivatives, which is uniquely given by coordinates $(u,\S)$ on the
phase plane.  Knowing this vector field, the solutions of the
equations are simply integral curves obtained by following the
vector field starting from any initial condition.  A special role
is played by the zeroes of this vector field, \ie, stable solutions
or fixed points of the differential equations, and by nullclines,
which are the lines along which one or the other derivative vanishes.

The nullclines of $u$ are the lines $\eta_0$, which corresponds to
$u=0$, and $\eta_*$, which is $\S=c_s^2-3$.  The nullcline $\eta_0$
happens to be an integral curve, since the vector field is everywhere
tangent to this line.  The nullclines of $\S$ are the hyperbolae
$\S^2+\B\S[u-(1+c_s^2)/\B]=\A$. This has two branches, $\sigma_-$ has
$\S<0$ and is asymptotic to the $u$-axis at $u=-\infty$, $\sigma_+$
has $\S>0$ and is asymptotic to the $u$-axis at $u=\infty$.

These two sets of nullclines cross at three fixed points--- $P_+=(u_+,\S_+)$
is the intersection of $\eta_0$ and $\sigma_+$, $P_-=(u_-,\S_-)$, which
is the intersection of $\eta_0$ and $\sigma_-$ and $P_*=(u_*,\S_*)$ which lies
on $\eta_*$ and $\sigma_-$, and
\beqa
\nonumber
   u_\pm &=& 0,\qquad
   \S_\pm = \frac12\left[1+c_s^2 \pm \sqrt{4\A + (1+c_s^2)^2}\right],\\
   u_* &=& \frac1{\B} \left[\frac{\A}{c_s^2-3}+4\right],\qquad
   \S_* = c_s^2-3.
\label{fixedpt}\eeqa
These features are shown in Figure \ref{fg.flows}

In a small interval around any fixed point $(u_f,\S_f)$ one can linearize
the equations to get
\beq
   \partial_\theta\left(\matrix{u-u_f\cr\S-\S_f}\right) =
        \M(u_f,\S_f)\left(\matrix{u-u_f\cr\S-\S_f}\right), \quad{\rm where}\quad
     \M(u,\S) = \left(\matrix{\frac14\left[\S+3-c_s^2\right] & u/4\cr
     -\B\S & -2\S + (1+c_s^2) - \B u}\right).
\label{linfluid}\eeq
At $P_\pm$ one of the off-diagonal components vanishes as a result
of which one can write down the eigenvalues by inspection. $P_+$
is a hyperbolic fixed point.  $P_-$ is a repulsive fixed point for
a Boltzmann fluid, but changes into a hyperbolic point for large enough
$\A$. One also finds that $P_*$ is a hyperbolic point for a Boltzmann
fluid but changes into a repulsive fixed point for sufficiently large $\A$. The
reason for this is not hard to understand. At $\A=4(3-c_s^2)$ the
points $P_-$ and $P_*$ are coincident and exchange character, leading
to a saddle-node bifurcation at this value of $\A$.

\subsubsection{The unstable manifold of $P_+$: a physically relevant solution}

 \begin{figure}[htb]\begin{center}
    \scalebox{0.5}{\includegraphics{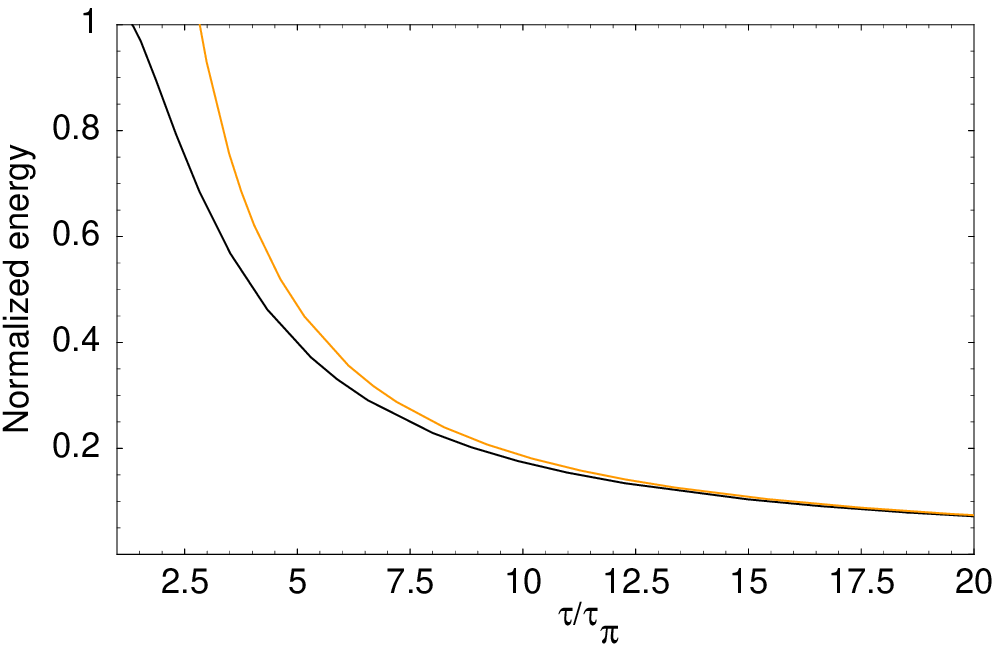}}
    \scalebox{0.5}{\includegraphics{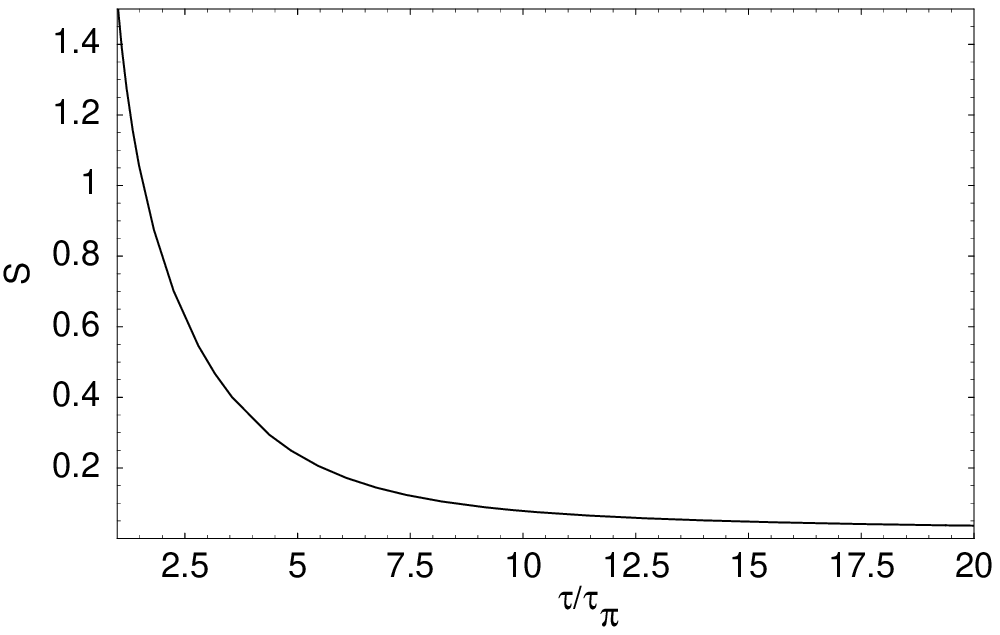}}
    \scalebox{0.5}{\includegraphics{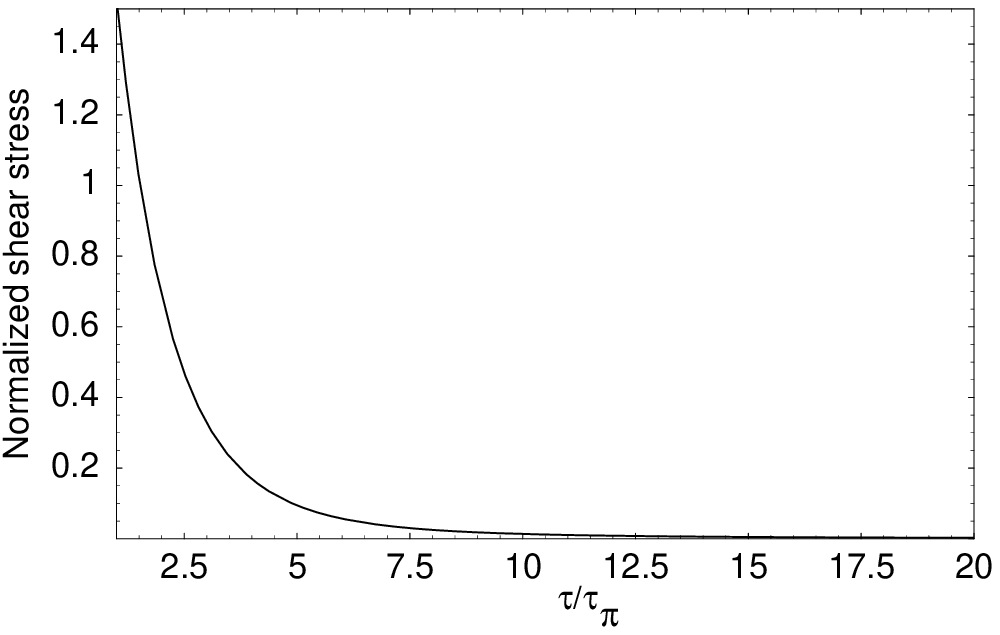}}
    \end{center}
   \caption{(Color online) The solutions corresponding to the unstable manifold for
   the normalized energy, $\epsilon/\epsilon_0=u(\theta)^4\mathrm e^{-4\theta}$, the quantity
   $\S$, and the normalized shear stress,
   $\S u(\theta)^4\mathrm e^{-4\theta}$, plotted against $\tau/\tau_\pi$.
   In the figure for the normalized energy, the curve in yellow shows the
   Bjorken solution normalized to match the exact numerical solution at
   large $\tau/\tau_\pi$.}
 \label{fg.unstable}\end{figure}

We concentrate on the physical flows, \ie, those originating on the
half-line $u=1$ and having $\S\ge0$. Since the vector $\V$ points into
the part of the plane bounded by $u=1$ and $\S=0$, one sees that physical
flows remain in this portion of the phase plane. Also, since there are
no fixed points in this part of the plane, one finds that flows cover
the whole quadrant.  Using Bendixson's theorems \cite{kbo}, one
can prove the intuitively obvious fact that there are no periodic or
quasi-periodic solutions starting from these initial conditions.

The main organizing element behind the physical flows is a special
solution, $\U$, \ie, a curve in the phase plane, called the unstable
manifold of $P_+$. This is the solution with initial conditions in
the infinitesimal neighbourhood of $P_+$, the explicit numerical
solution for which is exhibited in Figure \ref{fg.unstable}. Since
$P_+$ is a hyperbolic fixed point, all solutions starting from initial
conditions above $\U$ are attracted to it from above, and
those starting below it are attracted towards it from below. Clearly,
then, an important element of the analysis is $\S_\U$, the point at
which $\U$ intersects the line of physical initial conditions $u=1$.

Two varieties of stability in the solutions should be noted. At late
times the flows are stable against changes in initial conditions,
since they are always attracted to $\U$.  One useful consequence is
that fairly crude PDE solvers suffice to integrate eqs.\ (\ref{bfluid}).
There is a deeper level of stability, called structural stability, arising
from the fact that $P_+$ does not collide with any of the other fixed
points on changing the parameters $c_s^2$, $\A$, and $\B$. This means
that for all values of these parameters, the nature of physical flows is
determined by the corresponding solution for $\U$. The usefulness of
structural stability is that extraction of parameters from data becomes
particularly simple.

The first step to solving for the flows is to determine $\U$. In
the neighbourhood of $P_+$ it coincides with the eigenvector
corresponding to the positive eigenvalue of $\M(u_+,\S_+)$. Using eqs.\
(\ref{fixedpt}, \ref{linfluid}), it is easy to see that this is the line
$4\B\S_+u=(1+5c_s^2-9\S_+)(\S-\S_+)$. Using $\A=8/27$, $\B=1$ and $c_s^2=1/3$
gives $\S_\U=0.975913$ in this linear approximation.  The numerical
continuation of the straight line is easy, since any initial condition
close to $P_+$ will quickly settle down to $\U$. Such a construction
using the parameter values above gives $\S_\U=1.02545$, showing that
the slope of $\U$ decreases marginally beyond the linear analysis.

Solving for the distant part of $\U$ is equivalent to finding the
physical flows at long times. For this, we examine $u\gg1$. From the
phase space analysis we see that $\S$ decreases as $u$ increases when both
are in the physical region. Hence, in the second of eqs.\ (\ref{bfluid})
we can neglect the term in $\S^2$ with respect to the others. Also, the
term in $(1+c_s^2)\S$ can be neglected with respect to the remaining two
terms. Then the system of coupled equations to be solved is---
\beq
   \partial_\theta\left(\matrix{u\cr\S}\right) = \left(\matrix{
         u\left[\S+3-c_s^2\right]/4 \cr\A-\B u\S}\right).
\label{largetime}\eeq
Initial conditions $u_0$ and $\S_0$ are imposed at $\theta=0$ in order
to match this asymptotic solution with the full solution. Unfortunately,
even this equation is too complicated for an analytic solution.

We do not need all solutions of eq.\ (\ref{largetime}), as it happens. Since
$\varphi<4/3$ for hydrodynamics to apply, we must have
$\S<4/3\chi=\A$. For the Boltzmann fluid, then, $\S<8/27$. Of course, this
does not restrict all physical trajectories to approach $\U$ from below;
trajectories can approach this curve from above, but they correspond to
a different $\B$.  An explicit asymptotic form is easy to write down
when $\S\ll3$. Then $\S$ can be neglected in the equation for $u$,
and one gets---
\beqa
\nonumber
   u(\tau) &=& 
        u_0\left(\frac\tau{\tau_0}\right)^{(3-c_s^2)/4}
    \quad{\rm implying}\quad
   \epsilon(\tau) = \epsilon_0\left(\frac{\tau_0}\tau\right)^{1+c_s^2},\\
   \S(\tau) &=& \exp\left[-\frac{4\B u(\tau)}{3-c_s^2}\right]
     \left\{\S_0 +\frac{4\A}{3-c_s^2}
        {\rm Ei}\left( \frac{4\B u(\tau)}{3-c_s^2}\right)\right\}.
\label{asymp}\eeqa
Since these forms are asymptotically valid,
the constants $\S_0$, $u_0$ and $\epsilon_0$ are free parameters which ensure
that the asymptotic solution matches the exact solution at large $\tau$.
The forms above are not to be extrapolated down to small $\tau$.
In this limit one recovers Bjorken scaling, \ie, the boost-invariant
solution of the ideal gas. Not only is this a satisfactory mathematical
result, it could also be of physical relevance, if late freezeout occurs.
For later reference we note that at late times one has
\beq
   \S(\tau) = \frac{\A}{\B u_0} \left(\frac{\tau_0}\tau\right)^{(3-c_s^2)/4},
\label{sasymp}\eeq
using the asymptotic expansion of the exponential integral. The solution
illustrated in Figure \ref{fg.unstable} shows that, as a result of
viscosity, the expansion is slower than Bjorken, so that the energy
density is diluted less rapidly. As a result, the initial energy density,
as inferred from an observed final energy density, is very much smaller
than the Bjorken estimate.

From the phase plane analysis, it is clear that 
physical flows with initial conditions 
lying below $\S_{\cal U}$ are attracted to $\cal U$ from
below. For such solutions $\S$ increases initially before decreasing. Such
solutions have been exhibited in \cite{muronga,baier}. Other initial
conditions for physical flows give rise to monotonically decreasing
solutions for $\S$. The solution exhibited in Figure \ref{fg.unstable}
is the critical solution, $\cal U$, which separates these two types of
solutions. The long time behaviour of all solutions is arbitrarily close
to $\cal U$.

\subsubsection{The ELNS approximation}

As discussed before, the ELNS limit of the equations can be obtained by
dropping the term in the derivative of $\pi_V$, and then introducing an
arbitrary scale of time called $\tau_\pi$. In this approximation, the
equations for the Boltzmann fluid become
\beq
   u_\theta = \frac14u\left[\S+3-c_s^2\right],\qquad{\rm and}\qquad
     \B u\S=\A.
\label{belns}\eeq
Substituting the expression for $\S$ obtained from the second equation
into the first, the equation can be easily integrated with the initial
condition $u(\theta=0)=1$, to give
\beq
   u(\theta) = \mathrm e^{\theta(1-B/4)}f(\theta),\quad
   \frac{\epsilon(\theta)}{\epsilon_0} = \mathrm e^{-B\theta} f^4(\theta),
   \quad{\rm where}\quad
   f(\theta) = 1 + \frac{\A}{\B(4-B)}\left(1-\mathrm e^{(B/4-1)\theta}\right),
\label{belnssol}\eeq
and $B=1+c_s^2$. This is in the form of the Bjorken solution modified
by a factor which goes to a constant at large $\theta$. As before,
the ELNS solution matches the IS solution at times much larger than the
intrinsic time scale $\tau_\pi$.

\subsubsection{Entropy production}

 \begin{figure}[htb]\begin{center}
    \scalebox{1.0}{\includegraphics{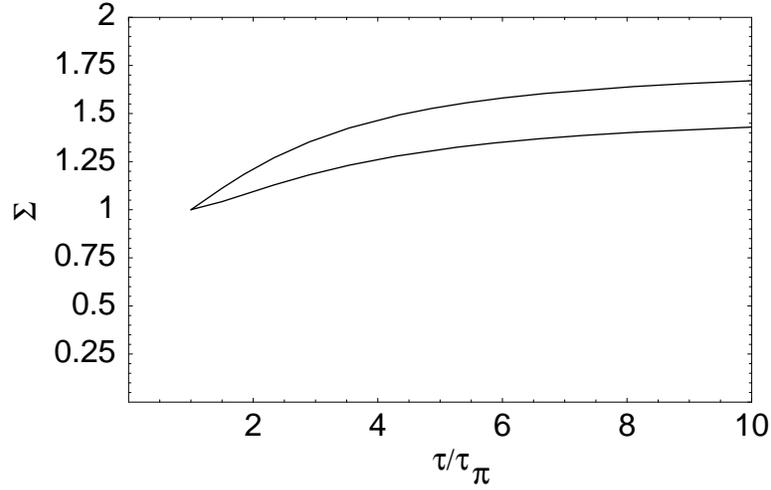}}
    \end{center}
   \caption{The normalized entropy $\Sigma$ as a function of $\tau/\tau_\pi$.
   Note the saturation at large $\tau$. The numerical solution is for
   $\A=8/27$ and $\B=1$. The upper curve is for $\S^0(0)=8/27$ and the
   lower for $\S^0(0)=2/27$.}
 \label{fg.entropy}\end{figure}

For a Boltzmann fluid the entropy density is proportional to $\sigma =
u^3\exp(-3\theta)$. One can then manipulate eq.\ (\ref{bfluid}) into the form
\beq
   \frac{d\sigma}{d\theta} = \frac34\sigma\left(\S-\frac43\right),
\label{bentropydensity}\eeq
where we have used the value $c_s^2=1/3$. Since $\S$ decreases with time,
it is clear that at sufficiently late time the factor $\S-4/3$ becomes
negative, and hence the entropy density must
decrease. The total entropy scales as $\Sigma=\sigma\exp\theta$, since the
spatial volume element picks up a scale factor of $\tau$ from the metric.
For this quantity we find the equation
\beq
   \frac{d\Sigma}{d\theta} = \frac34\Sigma\S,
\label{bentropy}\eeq
which is positive definite since the factors on the right hand side are
all positive. Hence the total entropy must increase. Using the expression
for $\S$ in eq.\ (\ref{asymp}), one finds that $\ln\Sigma$ can be written
in terms of Meijer-G functions. Using instead the asymptotic expansion of
$\S$ in eq.\ (\ref{sasymp}), one obtains the simpler long-time limit
\beq
   \Sigma(\tau)=\Sigma(\tau_0) \exp\left[ \frac{9\A}{8\B u_0}
       \left\{1-\left(\frac{\tau_0}\tau\right)^{2/3}\right\}\right].
\label{bbentropy}\eeq
In contrast to the simple fluid, where the entropy increases without
bound, the entropy of an expanding Boltzmann fluid reaches a finite
limit. The solution is shown in Figure \ref{fg.entropy}.

\subsection{Sound waves}

 \begin{figure}[htb]\begin{center}
    \scalebox{0.5}{\includegraphics{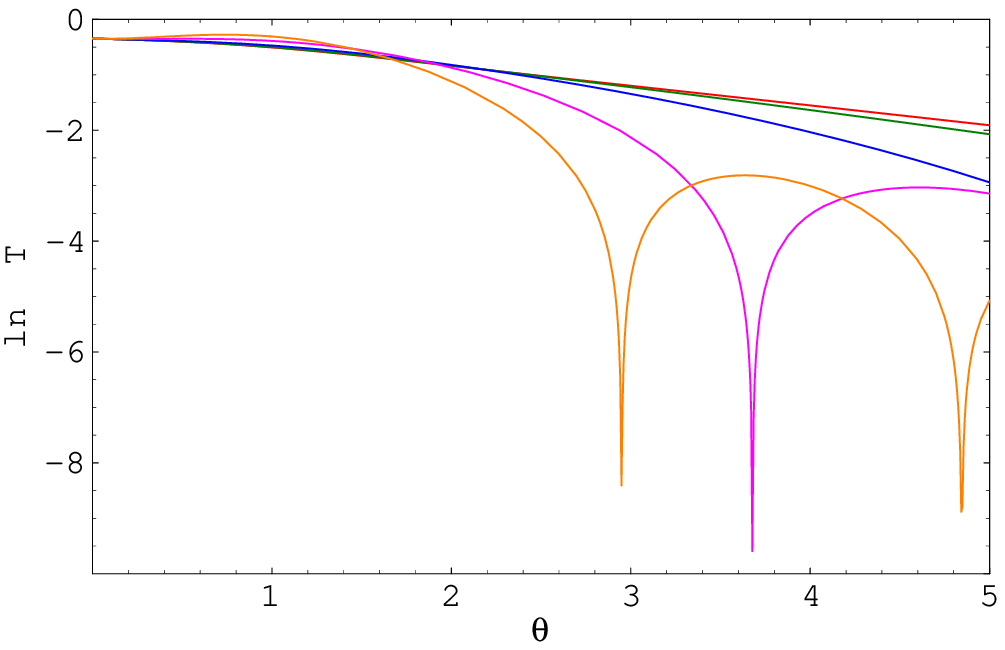}}
    \scalebox{0.5}{\includegraphics{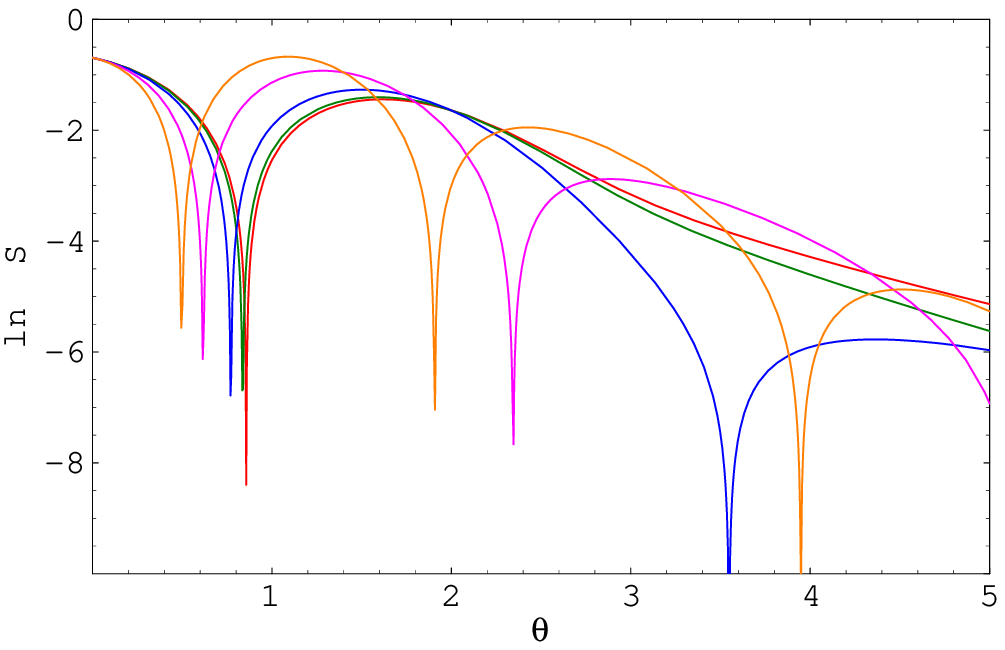}}
    \scalebox{0.5}{\includegraphics{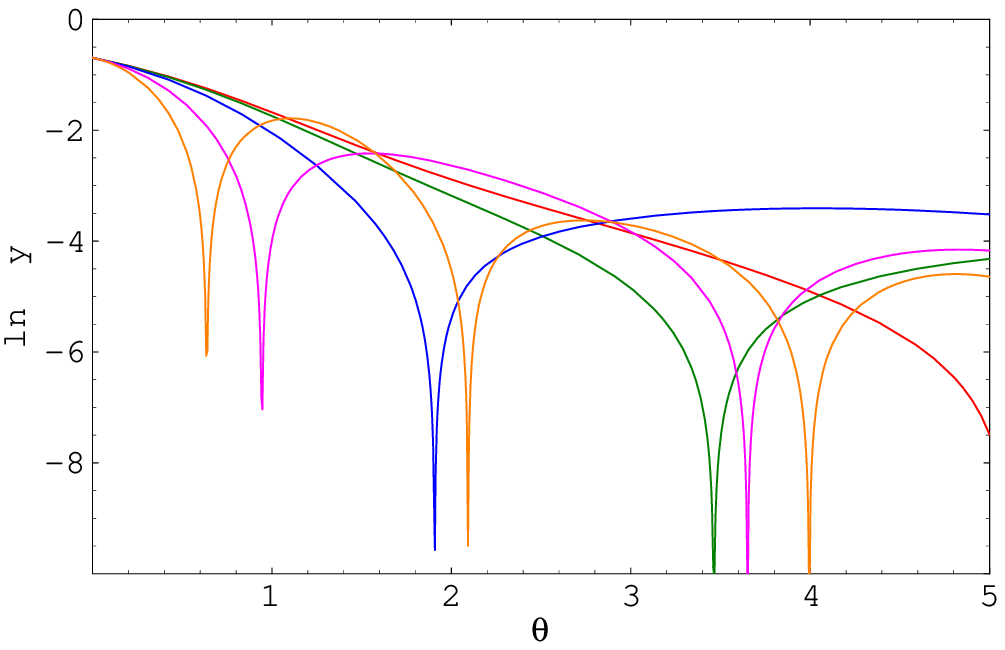}}
    \end{center}
   \caption{(Color online) Solutions of the linearized equations for fluctuations around
   the scaling solution in a Boltzmann fluid. The normalized temperature, $T$,
   the quantity, $\S$, and $y$ are shown for $k=1/4$
   (red), 1/2 (green), 1 (blue), 2 (purple) and 3 (orange) for $\S^0(0)=8/27$.
   The other initial conditions are $u^1(0)=1/\sqrt2$, $\S^1(0)=y^1(0)=1/2$.}
 \label{fg.boltzmannsound}\end{figure}

 \begin{figure}[htb]\begin{center}
    \scalebox{0.7}{\includegraphics{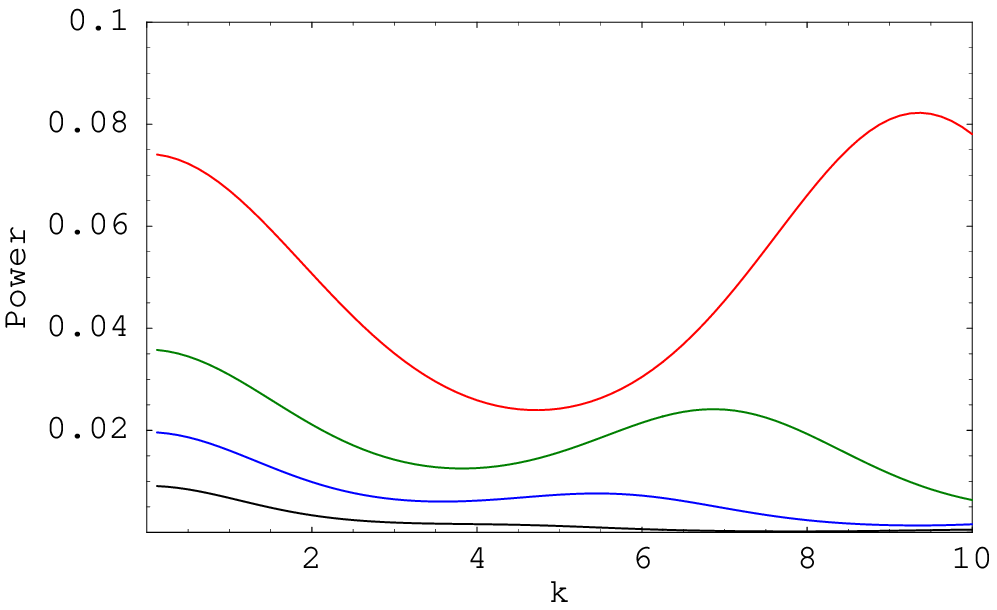}}
    \scalebox{0.7}{\includegraphics{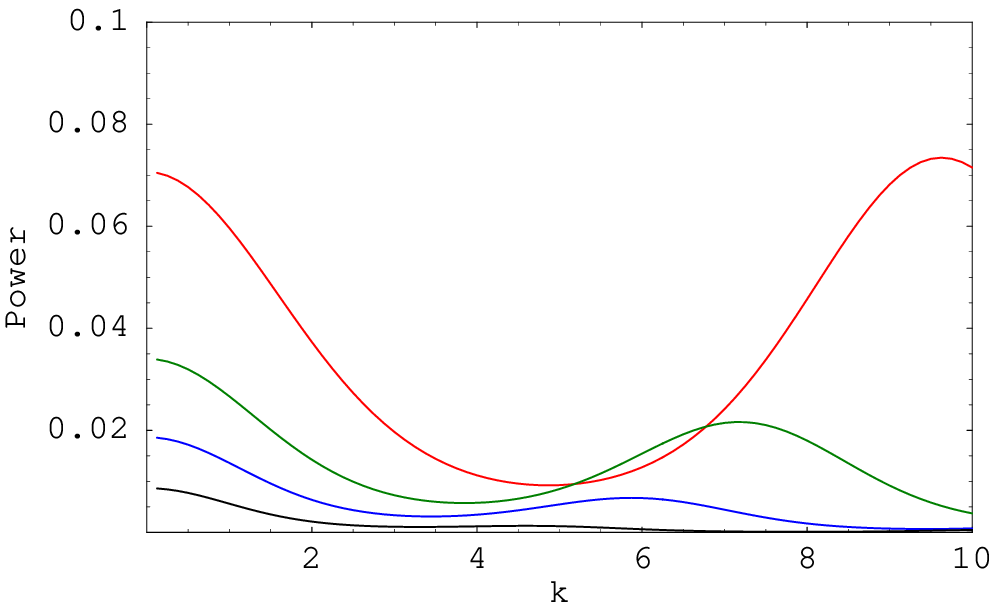}}
    \scalebox{0.7}{\includegraphics{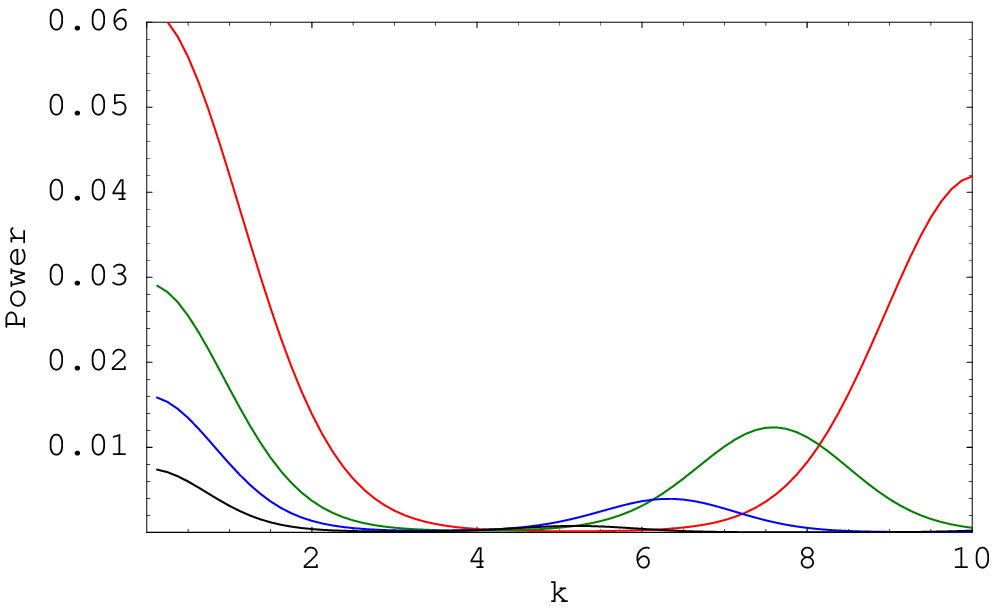}}
    \scalebox{0.7}{\includegraphics{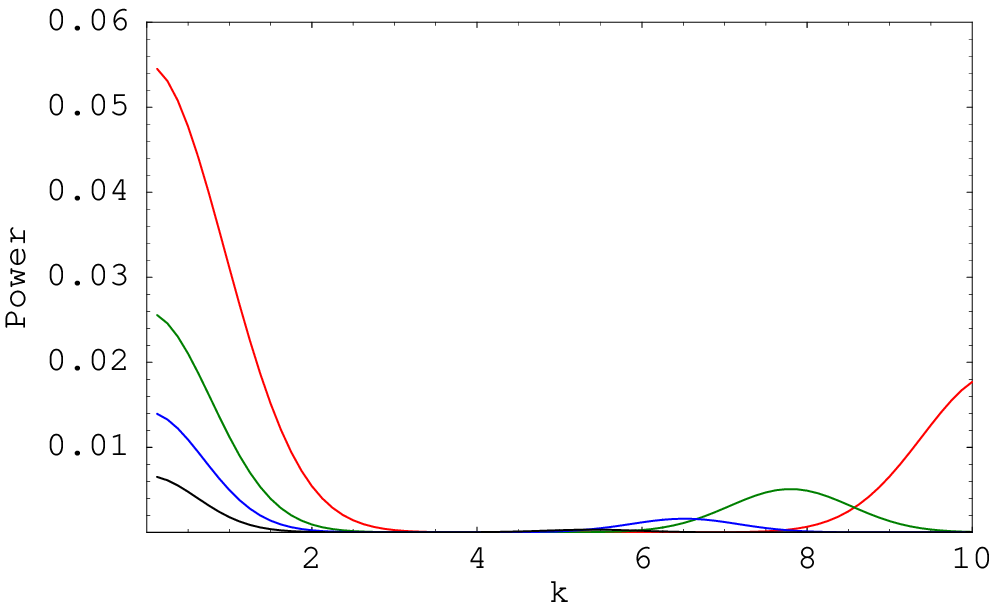}}
    \end{center}
   \caption{(Color online) The power spectrum of fluctuations after evolution from an
   initially flat spectrum at $\theta=0.8$ (red), 1 (green), 1.25 (blue)
   and 1.5 (black). The numerical solution is obtained for $c_s^2=1/3$,
   $\B=1$, $\A=8/27$ for four different values of $\S^0(0)$. The other
   initial conditions are $u^1(0)=-0.95$, $y^1(0)=-0.5$ and $\S^1(0)=0.3$.
   Note the difference in the scales.}
 \label{fg.power}\end{figure}

The linearized equations for fluctuations in a Boltzmann fluid are
\beqa
\nonumber
    u^1_\theta &=& \frac14(3-c_s^2+\S^0)u^1-\frac14(B-\S^0)u^0y^1_\eta
       +\frac14 u^0\S^1,\\
\nonumber
    y^1_\theta &=& \frac{4(c_s^2-\S^0)}{u^0(\S^0-B)}u^1_\eta
       +\frac{B(1-c_s^2)-\A+\S^0(\B u^0-1+c_s^2)}{\S^0-B} y^1
       -\frac1{\S^0-B}\S^1_\eta,\\
    \S^1_\theta &=&-\B\S^0 u^1-[(\S^0)^2-B\S^0-\A] y^1_\eta
       -(\B u^0+2\S^0-B) \S^1,
\label{boltzmannsound}\eeqa
where we have made the expansion $f(\theta,\eta)=f^0(\theta)+\Delta
f^1(\theta,\eta)$ where $f$ is any of $u$, $y$ and $\S$. As before,
$y^0=0$, and $u^0$ and $\S^0$ are obtained as the solution of eqs.\
(\ref{bfluid}), and we can Fourier transform in $\eta$ to examine the
evolution of each mode, $k$. Explicit solutions for $u^0$ and $\S^0$ can
be written only in the long-time limit, when these tend to the unstable
manifold $\U$.  However, some physically interesting phenomena are likely
to occur before this time. Hence, the method of choice is to numerically
solve eqs.\ (\ref{bfluid}, \ref{boltzmannsound}) together for the five
variables at the leading and first order in $\Delta$.  The initial
conditions for $u^0$ and $\S^0$ have been discussed before. Those for
$u^1$, $y^1$ and $\S^1$ can be chosen to lie between $-1$ and 1 in order
for $\Delta$ to give the right order of the magnitude of fluctuations.

The results obtained using $c_s^2=1/3$, $\A=8/27$, $\B=1$ are shown in
Figure \ref{fg.boltzmannsound}.
For large values of $k$, there are quasi-periodic
solutions. For smaller values of $k$ the solutions are overdamped. The
critical value, $k_0$, which separates damped and oscillatory solutions
depends on $\S^0(0)$. The solutions do not change qualitatively if $\B$
is changed by an order of magnitude. From the solutions displayed,
it is clear that a damping exponent can be extracted.

A quantity which encapsulates the physics, and is perhaps better suited
to making a connection with experiments is the power spectrum. In
Figure \ref{fg.power} we display the evolution of the power spectrum
starting from an initially flat spectrum. Qualitatively, the behaviour
is reminiscent of the simple fluid examined earlier. Evolution produces
peaks in the power spectrum. The positions of these peaks evolve with
time--- moving to smaller $k$ due to the redshifting discussed earlier.
The position and magnitude of the peaks depend very strongly on
initial conditions and $c_s$. For the relation between the power
spectrum studied here and the correlation function in \cite{romatschke},
see Section III.

The connection with ELNS hydrodynamics is made, as before, by dropping
the term in the time derivative of $\pi_V$. As we have mentioned before,
this results in the hyperbolic equations turning into a parabolic set.
As a result, fluctuations are diffusively damped, and do not turn into
propagating sound waves. This is seen in numerical solutions, and will
be dealt with more completely in the next section.

\section{A conformal fluid}

 \begin{figure}[thb]\begin{center}
    \scalebox{1.0}{\includegraphics{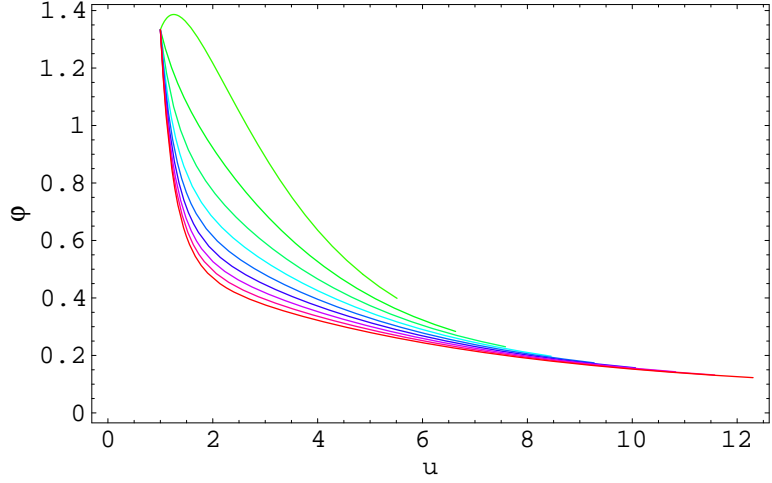}}
    \end{center}
   \caption{(Color online) Boost-invariant flow for conformal fluids in the plane of
      $\varphi^0$ and $u^0$, starting from the same initial conditions,
      with $\B=1$, but with varying $\chi$, \ie, $\A$. As $\A$ increases
      from $1$ to 9 in steps of $1$ ($\A$ increases from top to bottom),
      the long-time behaviour remains universal, but is approached faster.}
 \label{fg.binvconformal}\end{figure}

As discussed in Section II.A, a conformal fluid differs from a Boltzmann
fluid only in the $T$-independent value $\chi=\epsilon\tau_\pi/\vis
=3\pi a$, where $a=\tau_\pi T$. The hydrodynamics of such a conformal
fluid can then be taken over from that of the Boltzmann fluid with the
simple replacement $\A=4/3\chi=4/9\pi a$. In a strongly interacting
fluid one might expect $a=\tau_\pi T$ to be small, and hence $\A$ to be
large. The separation into boost-invariant and fluctuation equations
and the analysis of each, is very similar to the details presented in
Section V. For this reason, we do not repeat the material here, but only
point out the differences.

The boost-invariant flows follow eqs.\ (\ref{bfluid}). As discussed in
Section V.A, the global structure of the flows is determined by the
three fixed points $P_*$, $P_-$ and $P_+$.  When $\A$ is large enough,
the roles of the fixed points $P_-$ and $P_*$ are interchanged. However,
physical flows are governed by the unstable manifold of $P_+$ and
its behaviour remains unchanged. In particular, one may take over
the expressions for the long-time behaviour of the unstable manifold.

The boost-invariant flows, starting from the same initial
conditions, as $\chi$ changes, are shown in Figure \ref{fg.binvconformal}.
The late-time behaviour of the trajectories in $\varphi^0=\chi\S^0$ and
$u^0$ is independent of the value of $\A$. This is clear by using
eqs.\ (\ref{sasymp}) to write
\beq
   \varphi^0 = \frac{\A\chi}{\B u^0} = \frac4{3\B u^0}.
\label{universal}\eeq
This universality is a consequence of the structural stability of the
hydrodynamic equations for this class of fluids. From the figure one can
also see that for $\A\approx1$, $\varphi^0$ increases before decreasing.
Thermalization can be said to occur only when the solution enters the
physical domain $\varphi^0<4/3$ for the last time.

 \begin{figure}[bht]\begin{center}
    \scalebox{0.7}{\includegraphics{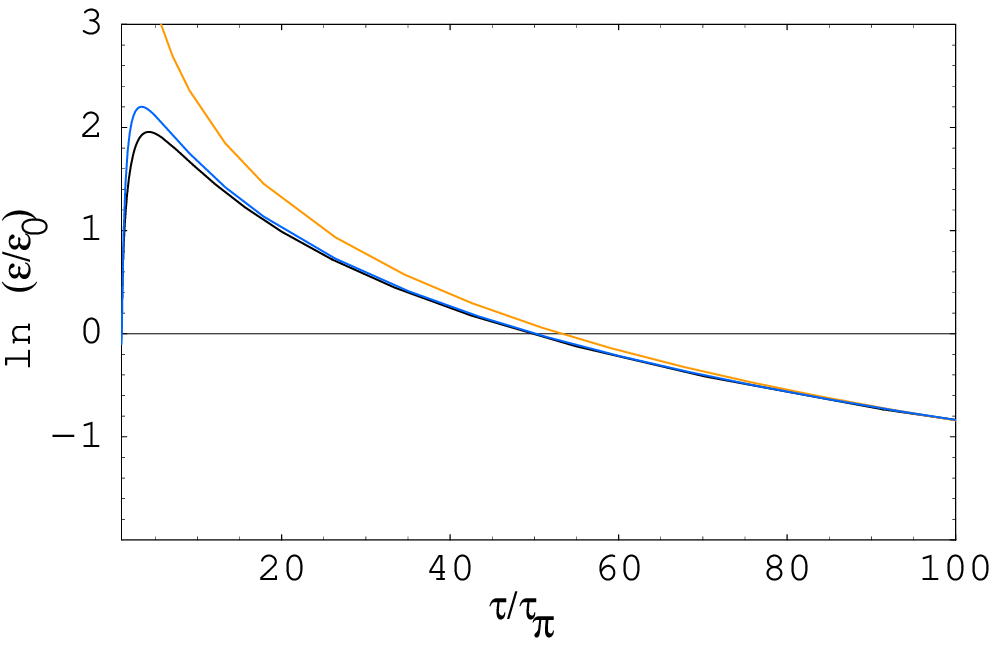}}
    \scalebox{0.7}{\includegraphics{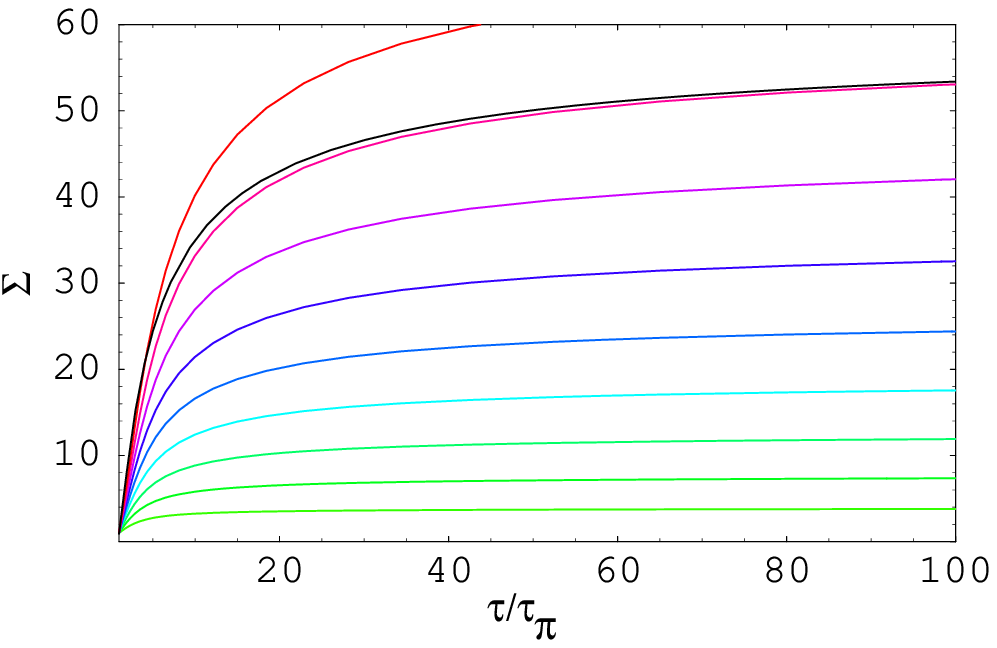}}
    \end{center}
   \caption{(Color online) Exact numerical solutions for (a) the energy density
     and (b) total entropy along the unstable manifold for conformal
     fluids. In (a) the line in black shows the exact solution for
     $\A=8$, the line in gold is a Bjorken solution matched to
     it asymptotically, and the line in blue is the corresponding
     solution in ELNS hydrodynamics matched asymptotically. In (b)
     the coloured lines show the time-evolution of $\Sigma$ as
     $\A$ varies from 1 to 9 in steps of 1, the larger the $\A$, the
     higher the curve. The line in black is the prediction from the
     ELNS solution matched to the energy density.}
 \label{fg.confelns}\end{figure}

From eq.\ (\ref{bbentropy}) it is clear that for all conformal fluids the
amount of entropy generated during the flow, proportional to $\Sigma$,
has a finite upper bound.
However, this bound increases exponentially with $\A$, starting from the
initial value $\Sigma(\tau_0)=1$, independent of $\A$.
Note that this means that at fixed value of $\vis/s$, the maximum entropy
production has exponential dependence on the inverse relaxation time,
$1/\tau_\pi$.

 \begin{figure}[bht]\begin{center}
    \scalebox{0.7}{\includegraphics{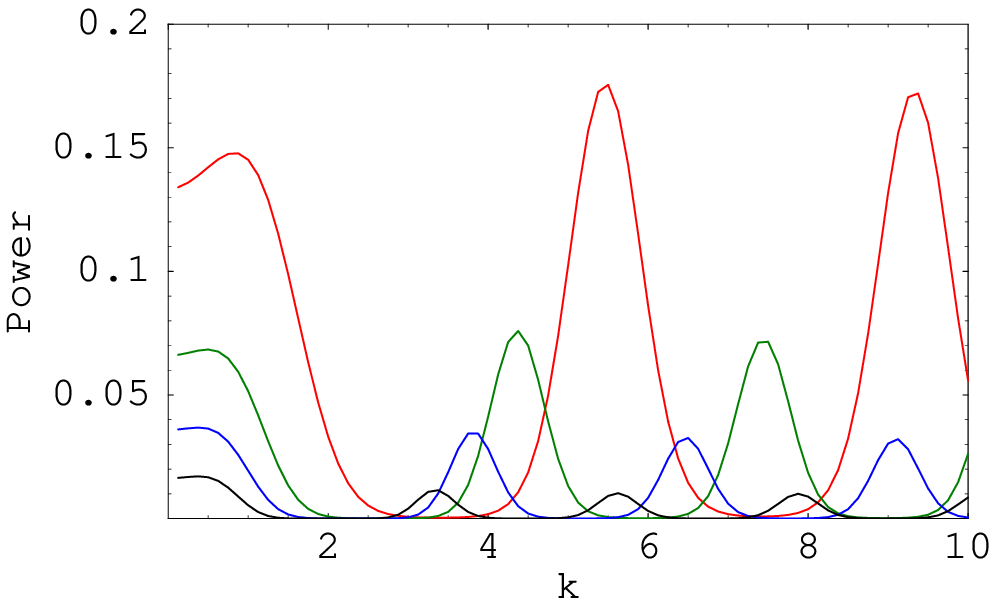}}
    \scalebox{0.7}{\includegraphics{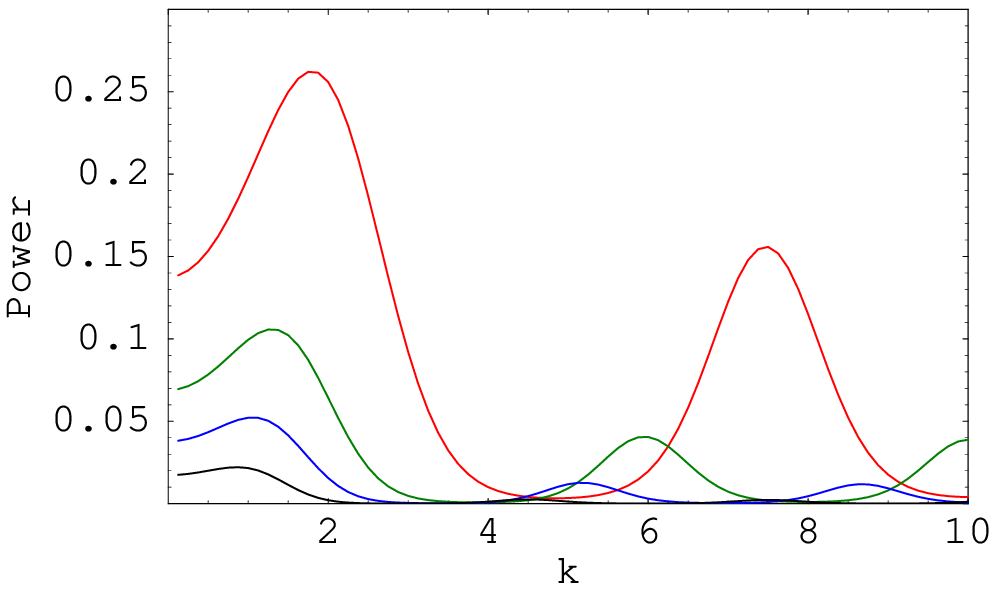}}
    \scalebox{0.7}{\includegraphics{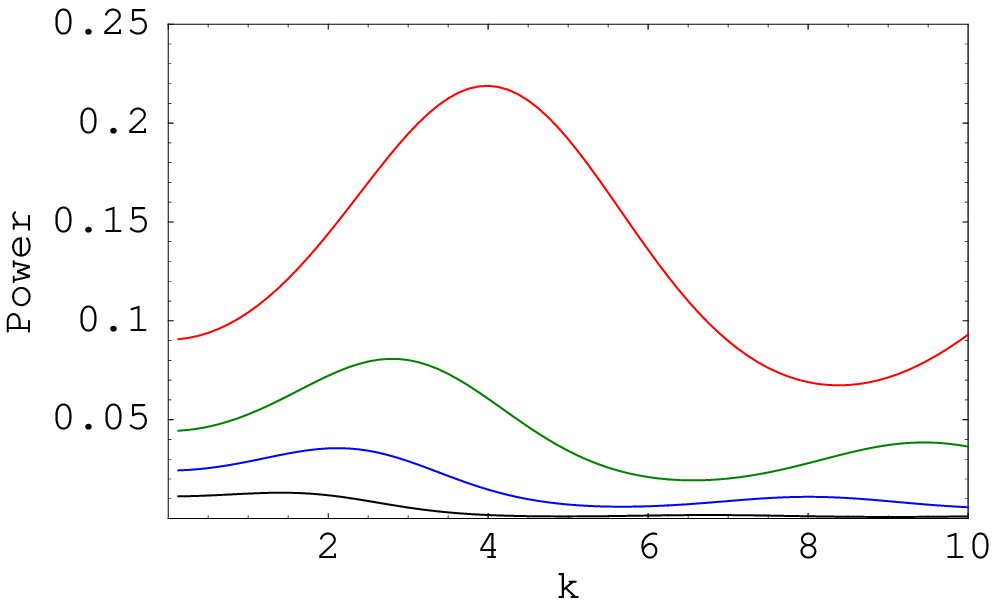}}
    \scalebox{0.7}{\includegraphics{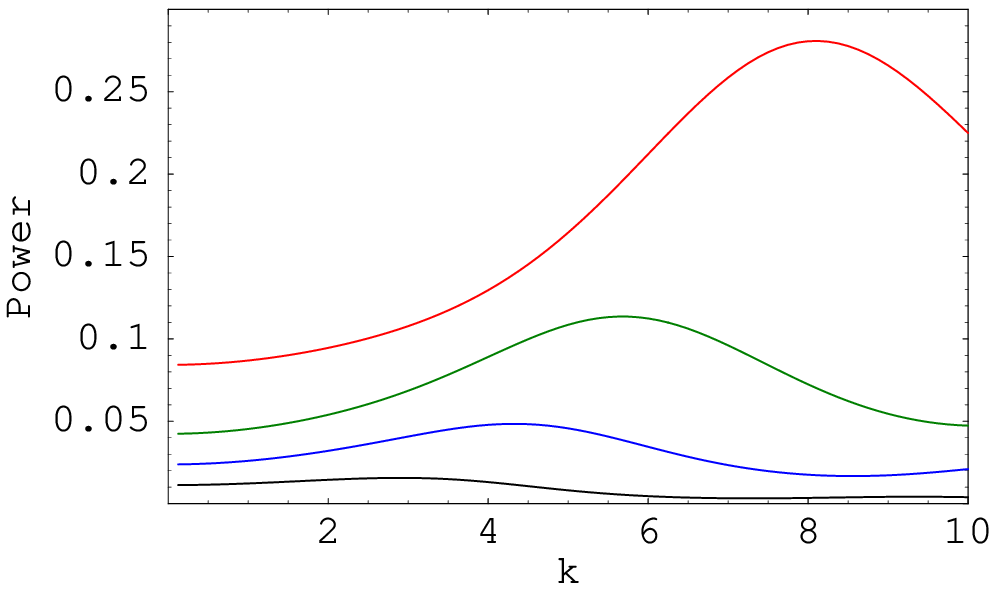}}
    \end{center}
   \caption{(Color online) The power spectrum for the energy density, at $\theta=0.8$ (red),
      1.05 (green), 1.25 (blue) and 1.5 (black), starting from a
      constant unit distribution with initial conditions $\S^0(0)={\rm min}(1,
      \A)$, $u^1(0)=-0.95$, $y^1(0)=-0.55$ and $\S^1(0)=0.30$ for all $k$.
      The successive figures are for (a) $\A=2$, (b) $\A=1$, (c) $\A=1/2$,
      (d) $\A=1/3$.}
 \label{fg.conformalsound}\end{figure}

Since the longitudinal flow can be continued to times of order $\tau_T$,
and $\tau_T$ is independent of any material property other than $c_s$,
for conformal fluids, the ratio $\tau_T/\tau_\pi$ could become large as
$\tau_\pi$ decreases. One might then expect that ELNS evolution should
suffice to describe the system. Figure \ref{fg.confelns} illustrates
several interesting points. First, the late time evolution of the energy
density can be approximated by an appropriately tuned Bjorken solution,
the tuning parameter being the initial energy density. However, as shown
in the figure, this gives a gross over-estimate of the energy density; in
the example, by more than two orders of magnitude. Second, a solution
of ELNS hydrodynamics can be tuned to reproduce the energy density at
late times. Once this tuning is performed, it also reproduces the total
entropy production at late times, and hence furnishes a far superior
description of the flow than the Bjorken solution could. Note, however,
that the ELNS solution has larger entropy production than the true IS
solution at initial times and that there are quantitative lacunae in this
approximation even at $\tau=40\tau_\pi$. With lower $\A$, we have seen
that these discrepancies are larger. Finally, we note that the maximum entropy
production occurs in the very early stages of the flow, and probes of
this stage of the expansion would best discriminate between different
values of $\tau_\pi$. A different approach to extracting $\tau_\pi$ is
advocated in \cite{dumitru}.

A possible discriminant between ELNS and IS hydrodynamics is the fate of
fluctuations around the scaling solution. An analysis of fluctuations can
be performed numerically using eqs.\ (\ref{boltzmannsound}), as before. 
At small $k$ all solutions are overdamped, as can be seen by investigating
the $k\to0$ limit, as before. At large $k$ the fluctuations develop into
damped propagating waves. The evolution of the power spectrum
of the fluctuations in energy density, starting from a uniform spectrum is
shown in Figure \ref{fg.conformalsound}.  For generic initial conditions,
increasing $\A$, \ie, decreasing $\tau_\pi$, seems to damp fluctuations
faster. However, the equations become stiff for $\A>2$ and the numerical
solutions are hard to extract for the interesting case of $N=4$ SYM theory
which yields $\A=8.69$.

However, in that case, we can take another approach.
The asymptotic solutions exhibited in eqs.\
(\ref{asymp}, \ref{sasymp}) are reliable for large $\theta$. One can investigate
the fate of fluctuations around the scaling solution at late times
by inserting the asymptotic formul{\ae} into eqs.\ (\ref{boltzmannsound}). The
asymptotic solutions can be written as $u^0=u_0\exp(p\theta)$ and
$\S^0=(\A/\B u_0)\exp(-p\theta)$, where $p=1-B/4$.
We expand eqs.\ (\ref{boltzmannsound}) in powers
of $z=\exp(p\theta)$, and retain all non-negative powers of $z$ in the
equations. This gives
\beqa
\nonumber
   &&\partial_\theta \left(\matrix{u^1\cr y^1\cr\S^1}\right) =
          M \left(\matrix{u^1\cr y^1\cr\S^1}\right),
      \qquad{\rm where}\qquad M = ikM_1+M_0,\\
     && \qquad\qquad M_1=\left(\matrix{
      0 & -\frac{Bu_0}4{\rm e}^{p\theta} & 0\cr
      0 & 0 & \frac1B\cr
      0 & \A & 0}\right),\qquad
     M_0=\left(\matrix{ p & 0 & \frac{u_0}4{\rm e}^{p\theta}\cr
      0 & -(1-c_s^2) & 0\cr
      0 & 0 & -u_0\B\mathrm e^{p\theta}}\right).
\label{conformalintheta}\eeqa
Note that the set $y^1$ and $\S^1$ can be solved independently of $u^1$,
and this last variable is then driven by the others.

For orientation, let us examine some analytical approximations first.
As $k\to0$, one may set $M=M_0$. The equations are then exactly
solvable, and yield
\beqa
\nonumber
   \S^1(\tau) &=& \S^1_0 \exp\left[\frac{u_0\B}p\left\{1-\left(
                        \frac{\tau}{\tau_0}\right)^p\right\}\right],\\
\nonumber
   y^1(\tau) &=& y^1_0 \left(\frac{\tau_0}{\tau}\right)^{1-c_s^2},\\
   u^1(\tau) &=& \left(\frac{\tau}{\tau_0}\right)^p \left[ u^1_0
       - \frac{u_0\S^1_0}{4p} {\mathrm e}^{\B u_0/p}\left\{
            {\rm Ei}\left(-\frac{\B u_0}p\right)
           -{\rm Ei}\left(-\frac{\B u_0}p\left(\frac{\tau}{\tau_0}\right)^p
               \right)\right\}\right].
\label{zerofreq}\eeqa
All these expressions must be truncated at order $z^0$ by dropping all terms
of order $1/z$ or smaller, since the equations were obtained similarly.
At large $\tau$, the temperature fluctuation, $u^1\tau_0/\tau$ decreases
as $1/\tau^{1-p}$, $y^1$ decreases as $1/\tau^{1-c_s^2}$, and $\S^1$ decreases
exponentially.

 \begin{figure}[bht]\begin{center}
    \scalebox{0.7}{\includegraphics{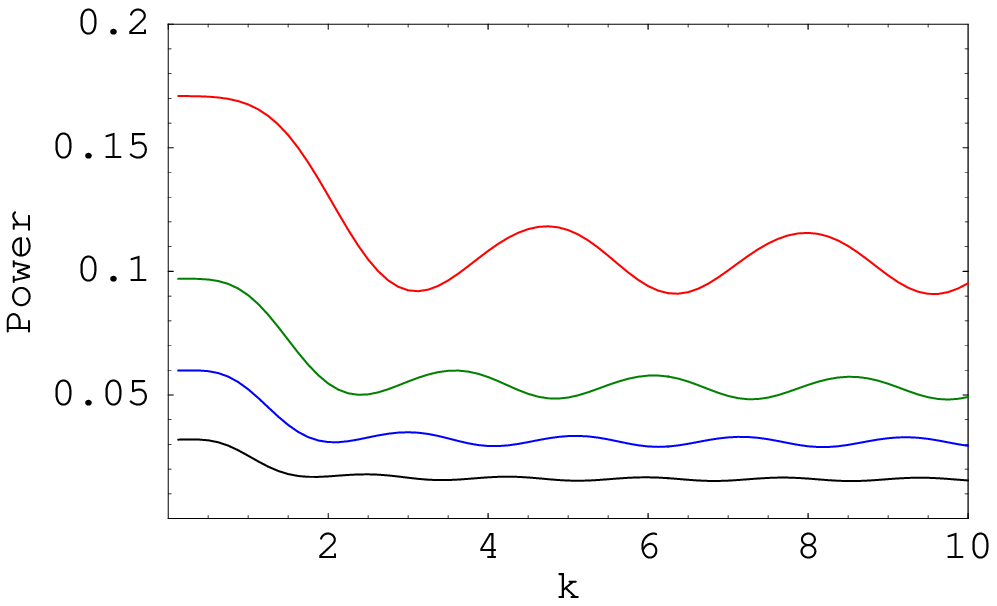}}
    \scalebox{0.7}{\includegraphics{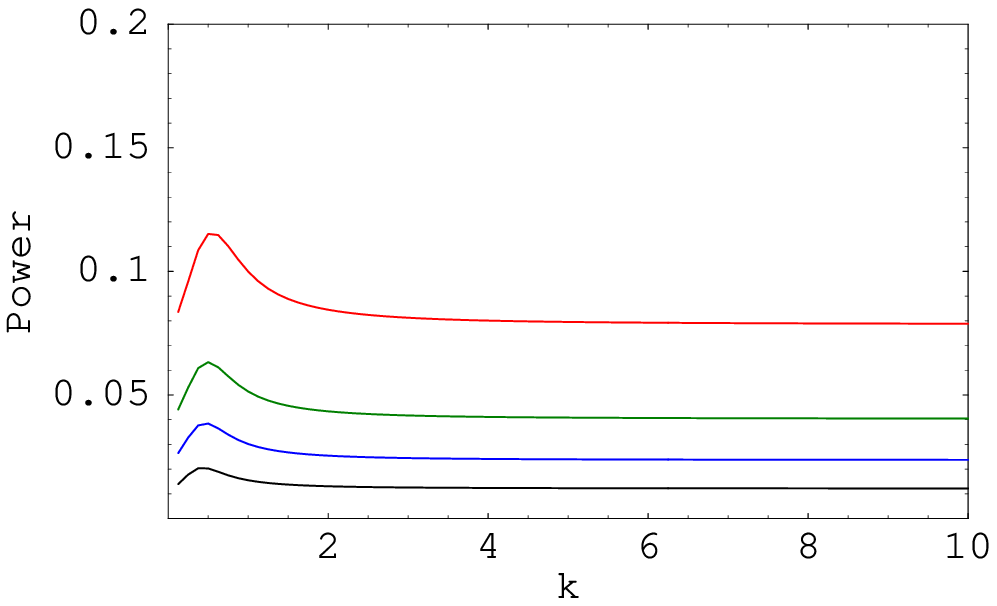}}
    \end{center}
   \caption{(Color online) The late-time power spectrum for the energy density, at
      $\theta=0.8$ (red), 1.05 (green), 1.25 (blue) and 1.5 (black),
      starting from a constant unit distribution in (a) IS hydrodynamics
      with $\A=8$ and initial conditions $u^1(0)=-0.95$, $y^1(0)=0.95$
      and $\S^1(0)=-0.75$ for all $k$ and (b) ELNS hydrodynamics with
      the same initial values of $u^1$ and $y^1$.}
 \label{fg.approxpower}\end{figure}

At large $k$, neglecting $M_0$, one finds that the solutions for $\S^1$ and
$y^1$ are oscillatory with frequencies $i\omega=\pm ik\sqrt{\A/B}$. The
solution for $u^1$ is entirely driven by $y_1$, and hence is oscillatory
with a growing amplitude. The growth exponent precisely matches the growth
exponent of $u^0$, hence $u^1/u^0$ is purely oscillatory in this approximation.
Damping arises with the inclusion of $M_0$. Then the equations are
not autonomous, and one could hesitate to treat the matrices $M_1$ and $M_0$
as time independent. However, by choosing $k$ to be large enough, one may
be able to treat $\exp(p\theta)$ as a constant over many periods of
oscillation. The simplest approach to computing the damping exponent then is
to treat the problem in perturbation theory. This is best done, as before, by
assuming that $i\omega=ik\lambda_1+\lambda_0+{\cal O}(1/k)$, introducing this
expansion into the characteristic equation for $M$ and then solving for
$\lambda_{0,1}$. As expected, $\lambda_1=\pm\sqrt(\A/B)$ and 0. The damping
exponent is $-\lambda_0$. We find that $y^1$ and $\S^1$ have the common
damping exponent $[1-c_s^2+u_0\mathrm e^{p\theta}]/2$. The temperature
fluctuations, $u^1/u^0$ are also damped.

The equations of ELNS hydrodynamics can be recovered from those of IS
hydrodynamics by the method which we have explained at length earlier. In
this case, the ELNS equations are obtained by first setting $\S^0=\A/\B
u^0$ (see eq.\ \ref{largetime}), an approximation which is equivalent
to the late-time solution in eq.\ (\ref{sasymp}), and reducing the last
of eqs.\ (\ref{conformalintheta}) through the further identification
$\S^1=\A y^1_\eta/\B$. Then the equation for
$y^1$ turns into a diffusion equation, which has a completely different
character from three equations of IS hydrodynamics which together give
rise to wave propagation.
The qualitative difference between the two kinds of hydrodynamics is
very clear from the evolution of the power spectrum of temperature
fluctuations, as shown in Figure \ref{fg.approxpower}. The propagation
of damped sound waves in IS hydrodynamics is signalled by the formation
of beats, and its absence in the case of ELNS signals the diffusive
nature of the dynamics. 

\section{Conclusions}

In this paper we examined freely expanding fluids using causal viscous
hydrodynamics \cite{causal} in the longitudinal 1-dimensional approximation.
We chose coordinates
appropriate to a description of longitudinal flow, performed a tensor
analysis and reduced the tensor equations of hydrodynamics to three
scalar equations.  The hydrodynamic modes are described by three scalars,
the energy density, $\epsilon$, shear viscous part of the energy-momentum
tensor, $\pi_V$, and a parametrization of the velocity, $y$.  Details of
this procedure are given in Appendices A and B.

Material properties expected of the QCD fluid were discussed in Section
II.A, and three models of the fluid were put forward. We gave the name
``simple fluid'' to a model in which $c_s$, $\vis$ and $\tau_\pi$ are
temperature independent.  Other toy models, which go
by the names of Boltzmann fluid and conformal fluids, have $c_s^2=1/3$
and $\chi=\epsilon \tau_\pi/\vis$, both independent of the temperature.
Different such fluids are distinguished by the value of $\chi$. For the
same value of $\vis/s$, one can have any value for $\chi$; large values
of the latter corresponding to gaseous fluids and small values to more
liquid-like behaviour.

In Section II.B we performed an analysis of the symmetries of the
hydrodynamic equations and extracted dimensionless scaling variables
which are combinations of the hydrodynamic variables and material
properties. These express laws of physical similarity, and can be related,
in the non-relativistic limit, to the familiar dimensionless variables,
\ie, the Reynolds number, $\rey$, the Mach number, $M$, and the Knudsen
number, $K$. For Boltzmann and conformal fluids, the condition that
$\chi$ does not depend on the temperature implies the combination $K\rey/M$
is constant. We discussed how IS hydrodynamics provides a self-consistent
description of thermalization.

In scaling flow of an ideal fluid, the initial energy density entirely
goes into driving the expansion. This leads to the Bjorken expansion with
its characteristic power law behaviour, $\epsilon\propto1/\tau^B$ (where
$B=1+c_s^2$).  In contrast, in a viscous fluid, some part of this energy
density is dissipated, leading to entropy production. Viscous effects are
sufficiently strong in the simple fluid that the power law is modified
to $\epsilon\simeq1/\tau$ asymptotically (section IV.A). As a result,
the total entropy of the system increases without bound (section IV.A.2);
we find $\Sigma\propto\tau^{1/4}$ asymptotically.

We made a phase plane analysis of flows for conformal fluids, including
the Boltzmann fluid (see Section V.A.1). Our analysis shows that
the long-time behaviour of any physical flow is arbitrarily close to
a special solution of the hydrodynamic equations which we call the
unstable manifold, $\cal U$.  We investigated this solution in detail
(see Sections V.A.2 and V.A.4), in particular, the approach to the
Bjorken solution and rates of entropy production.

In the Boltzmann fluid, Bjorken scaling is recovered asymptotically
(Section V.A.2). However, the initially slower expansion means that
the energy density is diluted slower; Bjorken estimates of the initial
energy density, given the final, are therefore over-estimates.  The total
entropy is saturated reasonably early (section V.A.4). Such behaviour also
holds for conformal fluids (section VI). However, for fixed $\vis/s$, the
saturation value for the entropy depends exponentially on the relaxation
time $\tau_\pi$. The rate of growth of the entropy does not depend
uniquely on $\vis/s$, since the parameter $\tau_\pi$ also plays a role.

When the relaxation time for the shear stress, $\tau_\pi$ is very small
in units of $1/T$ (\ie, the fluid behaves like a liquid), then the late
time solution for a conformal fluid is reasonably well approximated
in ELNS hydrodynamics by tuning a free parameter. This free parameter
is equivalent to the initial energy density (section VI). After
performing such a fit, the entropy density can also be fitted provided
an appropriate unit of time is chosen. Hence, the extraction of initial
conditions and material properties from observed final data remain as
equivalently hard problems in ELNS and IS hydrodynamics. Furthermore,
the initial conditions inferred from a given final energy density in
the two kinds of dynamics differ by a large factor. Hence, in making
inferences about the system produced in heavy-ion collisions, one must
be careful to use the right type of hydrodynamics.

For an ideal fluid, the fluctuations around the scaling solution are
overdamped for spatial Fourier modes $k<k_0=(1-c_s^2)/2c_s$. For larger $k$,
the fluctuations become damped sound waves. The energy density in the
fluctuations is damped as a power law in $\tau$ (section III). These
qualitative features persist in IS hydrodynamics of viscous fluids
(sections IV.B, V.B and VI). When $k$ is small enough, the solutions
are overdamped. At larger $k$ damped sound waves are obtained. The
scaling solutions are therefore stable against small fluctuations. The
power spectra of fluctuations contain interesting structure, which,
if observable, could give information on fluid properties or initial
conditions.

Although the scaling solutions obtained using IS hydrodynamics can
be well approximated at asymptotically late times by the solutions of
ELNS hydrodynamics (sections IV.A.1, V.A.3, VI), fluctuations behave
completely differently in these two kinds of dynamics. In the latter,
there are no sound modes: all fluctuations are diffusively damped. In
IS dynamics, however, modes with large enough $k$ give rise to true
sound waves. 
These two kinds of behaviour are easy enough to distinguish
through power spectra of the energy density (Figures \ref{fg.conformalsound},
\ref{fg.approxpower}). 
Similar effects are
also to be seen in the fluctuations of $y$.
Note the rather precise analogy of acoustically
produced peaks in the power spectrum of the temperature fluctuations
in heavy-ion collisions with those in the blackbody radiation in the
early universe.

One object of great modern interest in heavy-ion collisions has been the
coupling between hard and soft particles, for example jet-quenching. An
effective theory description of this is to treat the soft particles
through fluid dynamics and describe the hard particle as forcing a
shock wave in this medium. The behaviour of acoustics is an important
ingredient in such an analysis. In view of the difference between ELNS
and IS hydrodynamics in their treatment of fluctuations, they could
potentially give rise to different predictions for such observables.

The purpose of this paper was to explore the kind of phenomena that arise
in IS causal viscous hydrodynamics, and find qualitative features which
are different from either ELNS hydrodynamics or ideal gases. The changes
in the scaling solution are features which distinguish IS hydrodynamics
from ideal gases. The propagation of fluctuations around these solutions
as sound waves gives a qualitative distinction between IS and ELNS
hydrodynamics, since these modes are diffusively damped in the latter
case, and do not propagate.

Comparison with data, and fits to initial conditions and material
properties are outside the scope of this paper. These questions require
a proper treatment of radial and anisotropic flows, as well as the
incorporation of bulk viscosity (due to its importance near and below
$T_c$). These questions are left to the future.

\appendix
\section{Tensor decompositions}\label{sc.tensor}

\begin{table}[bht]
\begin{tabular}{||c|c|cccc||}
\hline
  & $g_{\mu\nu}$ & $L^\lambda_\mu$ & $V^\lambda_\mu$ & $\Sigma^\lambda_\mu$ & $A^\lambda_\mu$ \cr
\hline
$L^{\mu\nu}$ & 1 & $L^{\nu\lambda}$ & 0 & 0 & $\frac12(A^{\nu\lambda}+{\cal A}^{\nu\lambda})$ \cr
$V^{\mu\nu}$ & 1 & & $V^{\nu\lambda}$ & 0 & $\frac12(A^{\nu\lambda}-{\cal A}^{\nu\lambda})$ \cr
$\Sigma^{\mu\nu}$ & 2 & & & $\Sigma^{\nu\lambda}$ & 0 \cr
$A^{\mu\nu}$ & 0 & & & & $-L^{\nu\lambda}-V^{\nu\lambda}$ \cr
\hline
\end{tabular}
\caption{Contractions of the symmetric basis tensors. The entries are the
 contractions of the tensors in the row and column. The contraction in the
 first column is the trace.}
\label{tb.contract}\end{table}

In longitudinal flow there are only two vectors intrinsic to the problem---
the timelike $u$ and the spacelike $v$. One could construct two more
spacelike vectors to complete a tetrad, but since these vectors are
completely arbitrary, no physics can depend on them. In order to express
the rank-two tensors which enter the hydrodynamic equations, we can only
use combinations of $u$ and $v$, and the metric tensor.

The symmetric rank-two tensors can be chosen to be the projections
$L^{\mu\nu}=u^\mu u^\nu$, $V^{\mu\nu}=-v^\mu v^\nu$, and $\Sigma^{\mu\nu}
= g^{\mu\nu} - L^{\mu\nu} - V^{\mu\nu}$.  In addition there is the
traceless symmetric tensor $A^{\mu\nu}=u^\mu v^\nu+u^\nu v^\mu$, and
the antisymmetric tensor ${\cal A}^{\mu\nu}=u^\mu v^\nu-u^\nu v^\mu$.
The double contraction of the antisymmetric and any of the four symmetric
tensors vanishes. For notational convenience we also define the spacelike
projection $\Delta^{\mu\nu}=g^{\mu\nu} - L^{\mu\nu} = \Sigma^{\mu\nu} +
V^{\mu\nu}$. Contractions of this with the other tensors can be worked
out using Table \ref{tb.contract}. Any rank-2 symmetric tensor, which
arises in consideration of longitudinal flow, can be expressed as a
linear combination of $L$, $V$ and $\Sigma$ (or alternatively, $L$, $V$
and $A$). Any similar rank-2 antisymmetric tensor can only be a scalar
multiple of $\cal A$.

The vorticity tensor is defined as
\beq
   \omega^{\mu\nu} = \Delta^{\mu\alpha} \Delta^{\nu\beta} \frac12
      \left(d_\beta u_\alpha - d_\alpha u_\beta\right),
\label{vorticity}\eeq
and is spacelike (orthogonal to $u$) and antisymmetric by
construction. Due to its antisymmetry, $\omega^{\mu\nu}$ has to be
proportional to $\cal A$.  However, $\cal A$ is not orthogonal to $u$,
so the only possible constant of proportionality is zero. In other words,
vorticity vanishes for longitudinal flow. Another way to understand this
is to note that $\omega^{\mu\nu}$, being an antisymmetric spacelike
tensor, is equivalent (technically, dual) to an axial vector which is
spacelike. Such a vector can be constructed by the three-dimensional
vector product of two vectors. For longitudinal flow, there is only
one spacelike vector $v$ which is intrinsic to the problem. Hence one
cannot construct an axial vector. A tedious proof of the vanishing of
the vorticity can also be given by direct manipulation of the definition.

In the presence of viscous shear but vanishing bulk viscosity, one of
the hydrodynamic variables is the dissipative part of stress tensor,
$\pi^{\mu\nu}$. The fact that it is symmetric can be derived from the
symmetry of the stress tensor. Since it expresses shear, it is orthogonal
to $u$.  It is traceless since we have assumed bulk viscosity to vanish.
Hence one can write uniquely,
\beq
   \pi^{\mu\nu} = \pi_V\left[V^{\mu\nu}-\frac12\Sigma^{\mu\nu}\right],
\label{pivdef}\eeq
\ie, for longitudinal flow, the shear tensor is equivalent to a single
scalar. Note that the tensor multiplying $\pi_V$ has the following
properties---
\beq
   V^{\mu\nu}-\frac12\Sigma^{\mu\nu}=\frac32V^{\mu\nu}-\frac12\Delta^{\mu\nu}
     \qquad{\rm and}\qquad
   [V^{\mu\nu}-\frac12\Sigma^{\mu\nu}]
      [V_{\mu\nu}-\frac12\Sigma_{\mu\nu}] = \frac32.
\label{thattensor}\eeq

We will need to consider derivatives of the basis tensors. This is
facilitated by considering first the derivatives of the unit vectors.
The condition $u^\mu u_\mu=1$ yields $u^\mu Du_\mu=0$, and therefore
the orthogonality of $u$ and $v$. Now $D^2 u^\mu$ can be written as
a linear combination of $u$ and $v$. It is easy to work out that
$D^2 u^\mu=S^2u^\mu + (DS) v^\mu$. Furthermore, one can show that
$D v^\mu=Su^\mu$. Two other special cases of interest for longitudinal
flow are $d_\mu v^\mu=S$ and $v^\mu\D u_\mu=-\Theta$. For later use
we also set down the actions of $D=u^\mu d_\mu$ and $\D=v^\mu d_\mu$
on any scalar field $f$---
\beq
   Df = f_\tau \cosh y+\frac1\tau f_\eta\sinh y \quad{\rm and}\quad
   \D f = f_\tau \sinh y+\frac1\tau f_\eta\cosh y.
\label{derivonscalar}\eeq

Using the orthogonality of $u$ and $v$, we find the derivative
\beq
   \nabla_\mu u^\mu = d_\mu u^\mu = \Theta,\qquad{\rm where}\qquad
   \nabla_\mu \equiv \Delta_\mu^\nu d_\nu.
\label{trdu}\eeq
Another derivative that appears repeatedly in the hydrodynamic equations
is---
\beq
   \left\langle\nabla_\mu u_\nu\right\rangle \equiv
   \Delta_\mu^\lambda d_\lambda u_\nu + \Delta_\nu^\lambda d_\lambda u_\mu
      - \frac23\Theta \Delta_{\mu\nu},
\label{nablasym}\eeq
where the notation $\langle\cdots\rangle$ denotes the traceless, symmetric
part. Since this tensor is also orthogonal to $u$, one should
be able to write (see eq.\ \ref{pivdef})
\beq
   \left\langle\nabla_\mu u_\nu\right\rangle = 
       {\cal D} \left[ V_{\mu\nu} - \frac12 \Sigma_{\mu\nu} \right],
\label{notrace}\eeq
where $\cal D$ is to be determined. Multiplying through by $V_{\mu\nu}$,
one immediately finds
\beq
   \left\langle\nabla_\mu u_\nu\right\rangle =
       -2\left(v^\lambda\D u_\lambda + \frac\Theta3\right)
     \left[ V_{\mu\nu} - \frac12 \Sigma_{\mu\nu} \right] =
       \frac43\Theta \left[ V_{\mu\nu} - \frac12 \Sigma_{\mu\nu} \right].
\label{symntrdu}\eeq
Now we are ready to write the derivatives of the basis tensors.

Since covariant derivatives of the metric tensor vanish, one has
\beq
   d_\lambda\Delta_{\mu\nu} = -(d_\lambda u_\mu)u_\nu-(d_\lambda u_\nu)u_\mu.
\eeq
We examine its projections parallel and orthogonal to $u$. One of the parallel
projections is
\beq
   D\Delta_{\mu\nu} = -SA_{\mu\nu}.
\label{deltap}\eeq
One of the orthogonal projections is
\beq
   \Delta^\mu_\sigma\nabla_\rho\Delta^{\sigma\rho} = 
      \Delta^\mu_\sigma\Delta^\lambda_\rho d_\lambda\Delta^{\sigma\rho}
     =0.
\label{deltao}\eeq

For $V_{\mu\nu}$ one has
\beq
   d_\lambda V_{\mu\nu} = -(d_\lambda v_\mu)v_\nu-(d_\lambda v_\nu)v_\mu.
\eeq
A parallel projection which we will use later is
\beq
   D V_{\mu\nu} = -(D v_\mu)v_\nu-(D v_\nu)v_\mu = -S A_{\mu\nu}.
\label{vp}\eeq
One of the orthogonal projections that we need is
\beq
   \Delta^\mu_\sigma\nabla_\rho V^{\sigma\rho} = [S-d_\rho v^\rho] v^\mu.
\label{vo}\eeq
Since $v_\mu v^\mu=-1$, one has $v^\mu\D v_\mu=0$, \ie, $\D v_\mu$ is
parallel to $u$. As a result, one finds $\D v_\mu=(u^\lambda\D v_\lambda)u_\mu$,
from which the last form of the derivative follows.

\section{The equations of longitudinal hydrodynamics}\label{sc.hydroeqs}

The dynamical equations are supplemented by the equation of
state, which provides a relation between $\epsilon$ and $p$, and
hence determines the entropy density $s=(\epsilon+p)/T$. Since the
the hydrodynamic equations are valid only for a fluid in local
thermodynamic equilibrium, or so close to it that linear response
theory works, one may use the equation of state to eliminate one of
$\epsilon$ and $p$ from hydrodynamics. A toy equation
of state that we shall use is
\beq
   p(T) = c_s^2 \epsilon(T),
\eeq
\ie, the speed of sound is independent of the temperature. The only
value of $c_s^2$ which is strictly temperature independent is $c_s^2=1/3$.
This is the appropriate value to use when the bulk viscosity has been
neglected, since both are consequences of conformal symmetry. Since we
use temperature independent $c_s^2$ in this paper, we use the above value
whenever numerical work is performed.

\subsection{The equation for energy}

The equation for energy in \cite{baier} is
\beq
   D\epsilon = -(\epsilon+p)\nabla_\mu u^\mu + \frac12\pi^{\mu\nu}
       \langle\nabla_\mu u_\nu\rangle.
\label{entropybaier}\eeq
The identity in eq.\ (\ref{thattensor}),
used along with eq.\ (\ref{symntrdu}) gives
\beq
   \pi^{\mu\nu}\langle\nabla_\mu u_\nu\rangle =
      2\pi_V \Theta.
\eeq
Then, using the relation in eq.\ (\ref{trdu}), and the equation of
state, one can write
\beq
   D\epsilon = -\left[(1+c_s^2)\epsilon - \pi_V\right]\Theta.
\label{energyeq}\eeq

\subsection{The momentum-balance equation}

The general form of the momentum-balance equation given in \cite{baier} is
\beq
   (\epsilon+p) Du^\mu = \nabla^\mu p - 
     \Delta^\mu_\sigma\nabla_\rho\pi^{\sigma\rho}
      + \pi^{\mu\sigma}Du_\sigma.
\eeq
Note that each term is orthogonal to $u$.
Using the definition of $v$ and the decomposition of the shear tensor
in eq.\ (\ref{pivdef}), we find that
\beq
   \pi^{\mu\sigma}Du_\sigma = S\pi^{\mu\sigma}v_\sigma = S\pi_V v^\mu.
\eeq
Using the derivatives of the basis tensors in eqs.\ (\ref{deltao},
\ref{vo}), we can write
\beq
   \Delta^\mu_\sigma\nabla_\rho\pi^{\sigma\rho} =
       \Delta^\mu_\sigma\Delta^\lambda_\rho (d_\lambda \pi_V)
      [V^{\sigma\rho}-\frac12\Sigma^{\sigma\rho}]
        - \frac32\pi_V v^\mu [d_\lambda v^\lambda-S].
\eeq
The last term drops out because $d_\lambda v^\lambda=S$.
Putting all this together, we can reduce the tensor equation to
\beq
   S(\epsilon+p)v^\mu = \nabla^\mu p - (d_\lambda\pi_V)
    [V^{\lambda\mu}-\frac12\Sigma^{\lambda\mu}] + S \pi_V v^\mu.
\eeq
Contracting with any spacelike tensor apart from $v$ would yield
only terms in the directional derivative along that vector. But for
longitudinal flow each such derivative is separately zero. Thus,
the only nontrivial equation is obtained by contracting the above
equation with $v$. This gives the entropy equation for longitudinal
flow,
\beq
   S(\epsilon+p-\pi_V)+\D(p-\pi_V) = 0.
\label{entropyeq}\eeq

\subsection{The equation for the shear tensor}

In \cite{baier} the equation for the shear tensor is given as
\beq
   \tau_\pi \Delta^\mu_\alpha\Delta^\nu_\beta D\pi^{\alpha\beta}
   + \pi^{\mu\nu} = \vis\langle\nabla^\mu u^\nu\rangle
   - 2\tau_\pi \pi^{\alpha(\mu}\omega^{\nu)}_\alpha,
\eeq
where $\tau_\pi$ is a relaxation time, $\vis$ is the coefficient of
shear viscosity, $\nabla^\mu=\Delta^{\mu\nu}d_\nu$, and
$\omega^{\alpha\beta}$ is the vorticity tensor. Since the vorticity
vanishes for longitudinal flow, we drop the last term.

Using eq.\ (\ref{symntrdu}) the term in the viscous coefficient becomes
\beq
   \vis\langle\nabla^\mu u^\nu\rangle 
      = \frac43\vis \Theta[V^{\mu\nu}-\frac12\Sigma^{\mu\nu}].
\label{visc}\eeq
For the first term, one can write
\beq
   \Delta^\mu_\alpha\Delta^\nu_\beta D\pi^{\alpha\beta}
      = (D\pi_V)[V^{\mu\nu}-\frac12\Sigma^{\mu\nu}],
\label{firstterm}\eeq
since the projectors are orthogonal to the derivatives of the
basis tensors in eqs.\ (\ref{deltap}, \ref{vp}).
The equation for shear then reduces to the scalar equation
\beq
   \tau_\pi (D\pi_V) + \pi_V = \frac43\vis\Theta.
\label{sheareq}\eeq

\section{Evolution of the power spectrum}

For the equations of sound we assume 
that a Fourier transformation decouples the individual Fourier modes. For
each mode, the ODEs to be solved can be written in the form
\beq
   \frac{d\mathbf x}{d\theta} = M \mathbf x,
\label{ode}\eeq
where $\mathbf x$ is a vector and $M$ a matrix. For ideal fluids, eqs.\
(\ref{idealsound}) can be written as a system of two coupled autonomous
equations, so that $M$ is a $2\times2$ matrix.  In the other cases (eqs.\
\ref{simplesetintheta}, \ref{boltzmannsound}
and its reduction in the ELNS limit) the equations are not autonomous,
\ie, the time variable appears explicitly in $M$.  However, the
solution of eq.\ (\ref{ode}) is straightforward and involves a matrix
exponential.  The asymptotic behaviour of the solution is controlled
by the eigenvalue with the largest real part.  We will assume that
all the eigenvalues of $M$ have negative real part (at all times, in
the non-autonomous case), so that at long times the solution decays.
We are interested in whether transients grow.

Consider the real positive quantity $P(\theta) = \mathbf x^\dag A
\mathbf x$ where $A$ is a fixed matrix independent of $\theta$. Let
${\cal M}=M^\dag A+AM$.  Then
\beq
   \frac{dP}{d\theta} = \mathbf x^\dag {\cal M} \mathbf x
     = \sum_i |x_i|^2 \lambda_i
    \qquad{\rm where}\qquad \mathbf x=\sum_i x_i\mathbf v_i,
\label{norm}\eeq
in terms of the normalized eigenvectors, $\mathbf v_i$, and eigenvalues,
$\lambda_i$, of $\cal M$. Since
$P$ is real, $A$ is Hermitean. As a result, $\cal M$ is also Hermitean,
and its eigenvalues are real. Transient growth in $P$ can take place for
some initial conditions if and only if $\cal M$ has at least one positive
eigenvalue. The largest growth in $P$ occurs when $\mathbf x$ is parallel
to the eigenvector of $\cal M$ with largest (positive) eigenvalue.

 \begin{figure}[htb]\begin{center}
    \scalebox{0.4}{\includegraphics{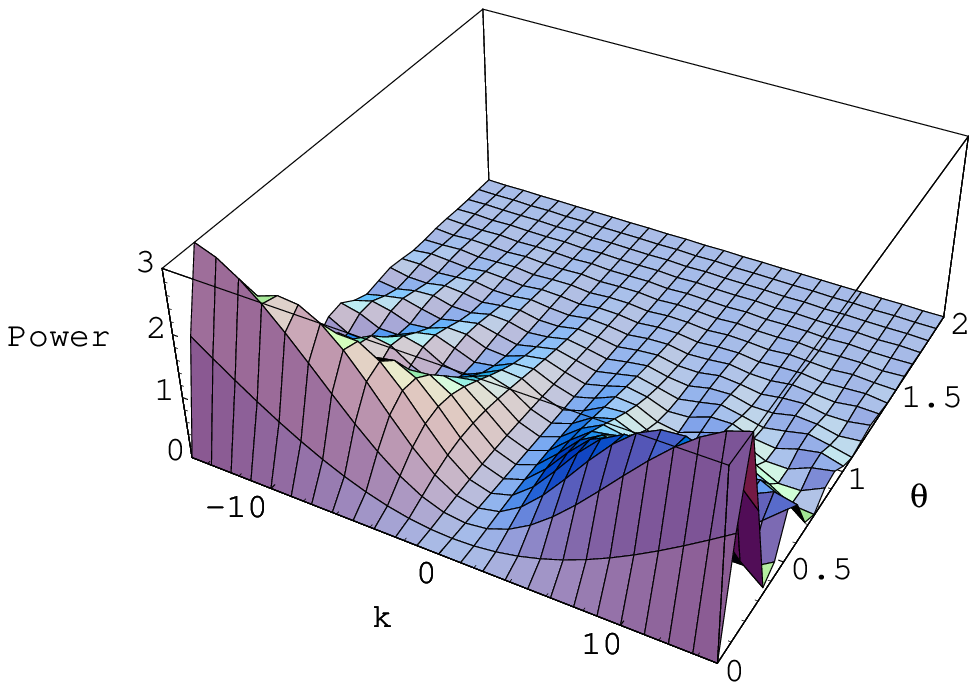}}
    \scalebox{0.4}{\includegraphics{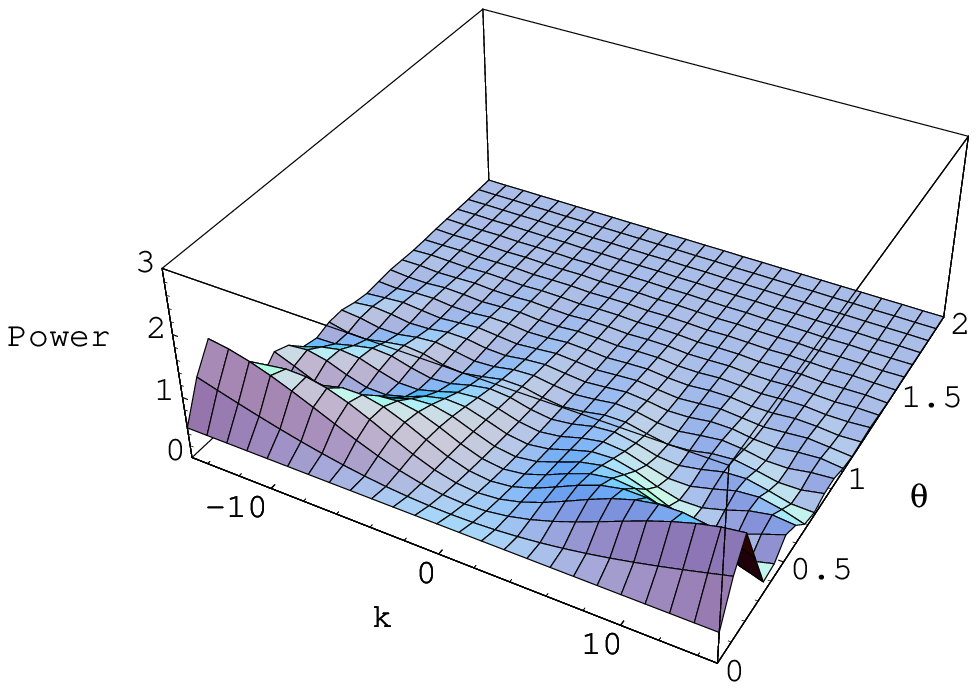}}
    \scalebox{0.4}{\includegraphics{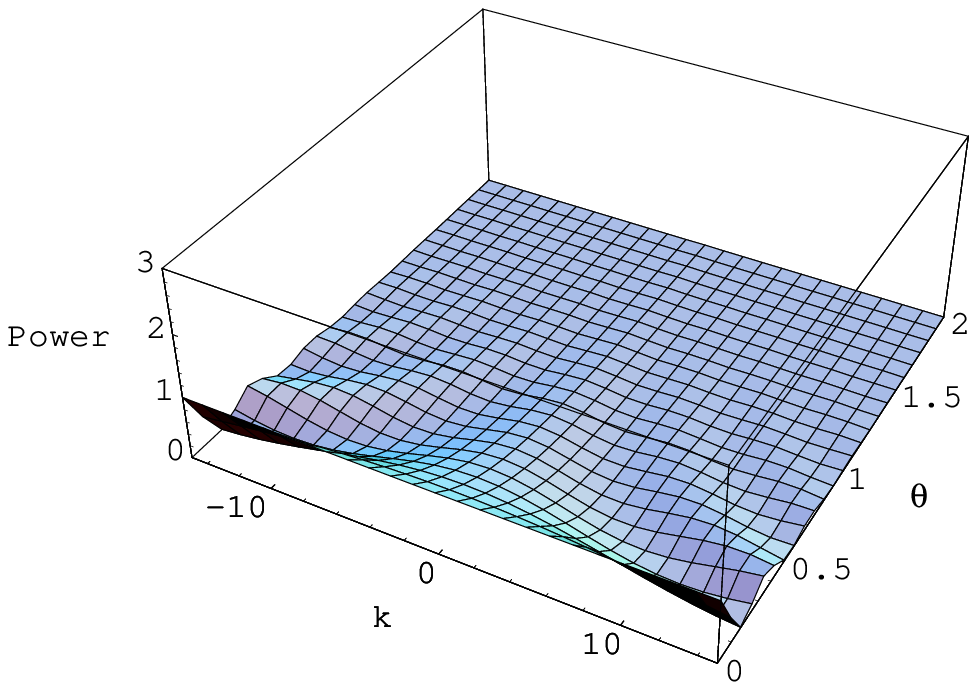}}
    \scalebox{0.4}{\includegraphics{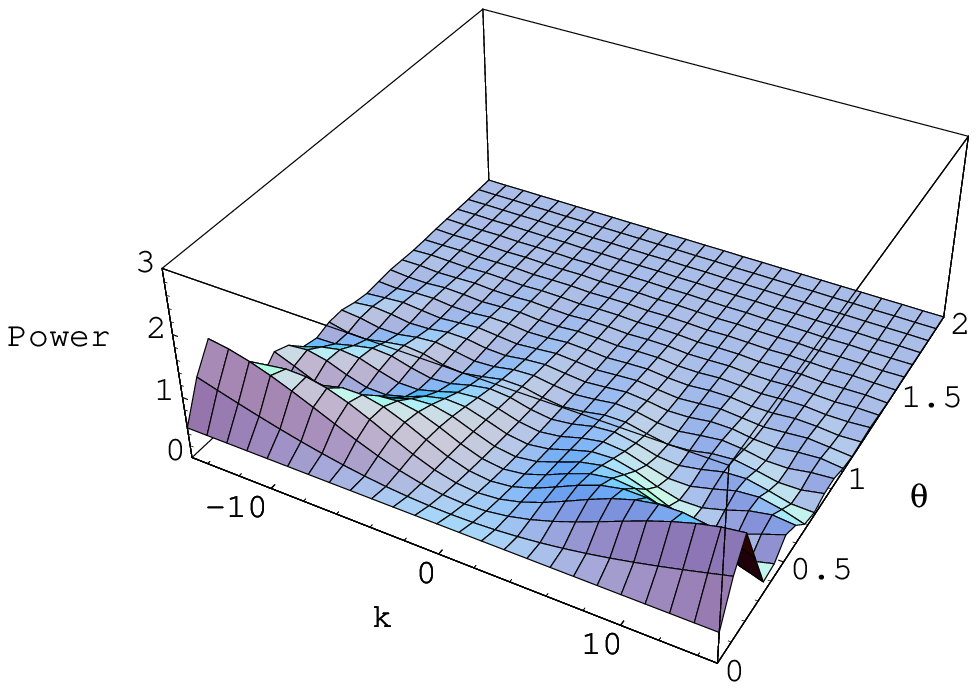}}
    \end{center}
   \caption{(Color online) The power spectrum of fluctuations in energy density
     for the ideal fluid as a function of $k$ and $\theta$, for
     $\beta=0$ and $\alpha=0$, $\pi/4$, $\pi/2$ and $3\pi/4$
     (from left to right). Note the lack of transient growth for $k=0$.}
 \label{fg.actualideal}\end{figure}

 \begin{figure}[htb]\begin{center}
    \scalebox{0.4}{\includegraphics{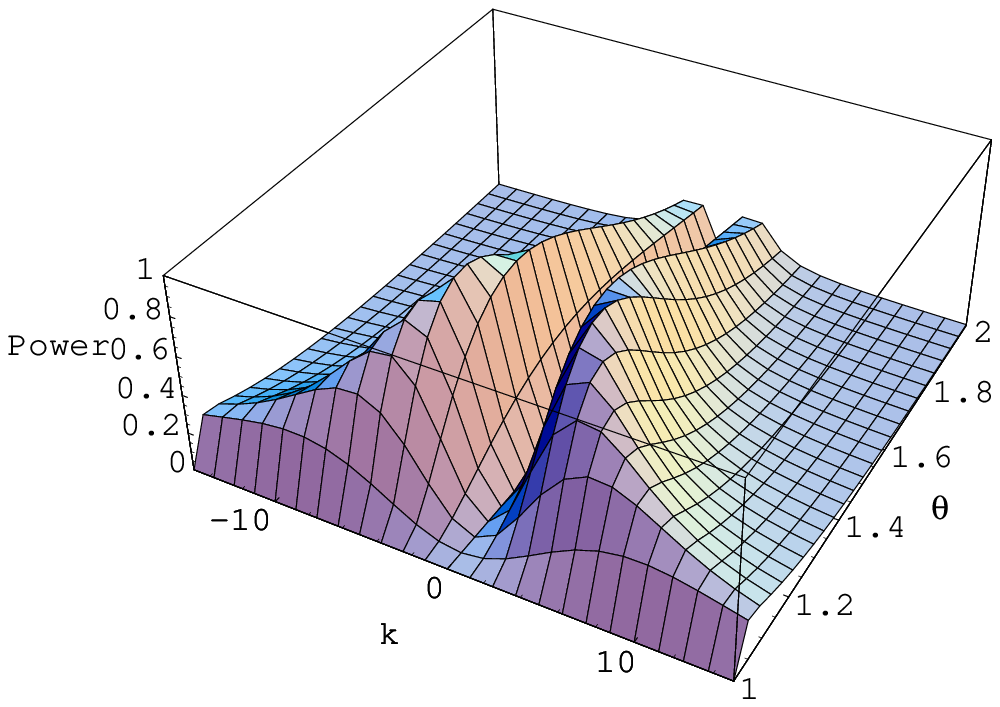}}
    \scalebox{0.4}{\includegraphics{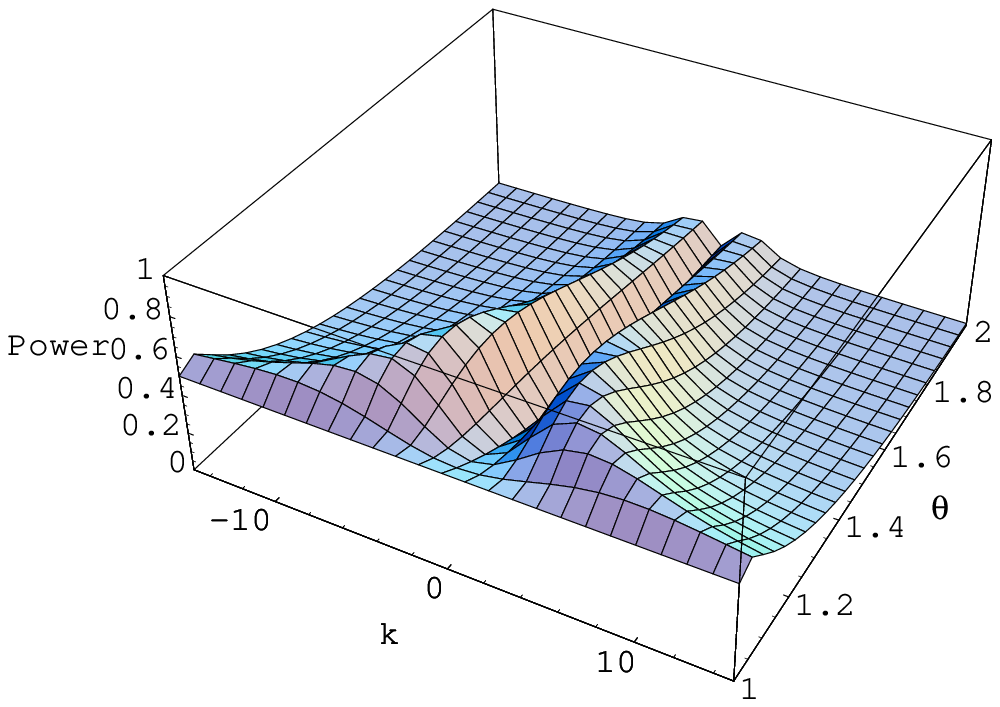}}
    \scalebox{0.4}{\includegraphics{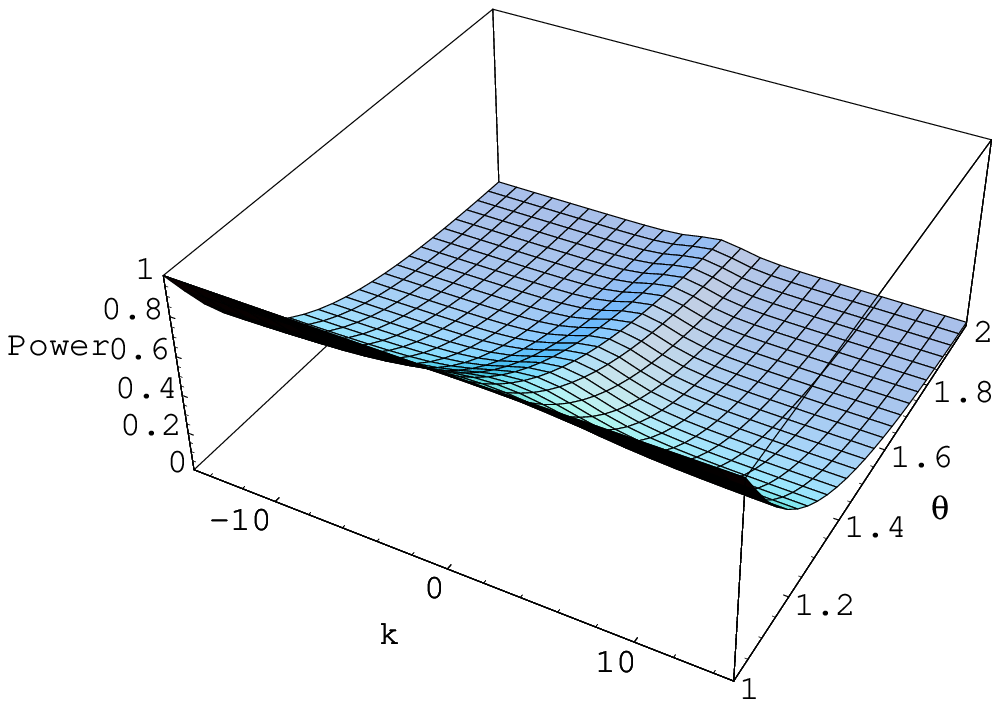}}
    \scalebox{0.4}{\includegraphics{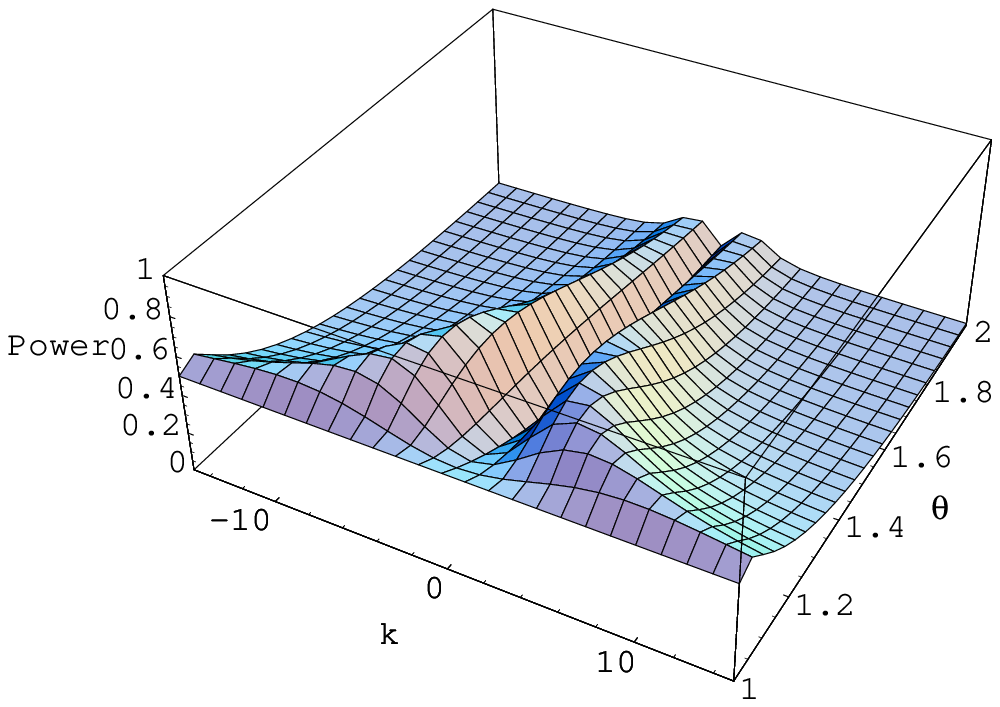}}
    \end{center}
   \caption{(Color online) The power spectrum of fluctuations in energy density
     for the simple fluid in ELNS hydrodynamics as a function of
     $k$ and $\theta$, for $\beta=0$ and $\alpha=0$, $\pi/4$,
     $\pi/2$ and $3\pi/4$ (from left to right). Note the lack of
     transient growth for $k=0$.}
 \label{fg.actualelnssimplesound}\end{figure}

When $A$ is chosen to be a multiple of the identity, then the
eigenvalues of $\cal M$ are multiples of the real parts of the
eigenvalues of $M$.
Then, since the multiplying constant is positive,
and we have assumed that the eigenvalues of $M$ have negative real parts, there
can be no growth in $P$. In all other cases, transient growth of $P$
is possible.

In our application, we choose $A$ to be a projection operator on the first
component of $\mathbf x$. As a result, $AM$ is just the top row of $M$
with other rows set to zero, and $\cal M$ is obtained by Hermitizing this.
Now the eigenvalues of $\cal M$ have no simple relation with those of $M$.
One can show that in general the rank of $\cal M$ is two, \ie, there are
two non-vanishing eigenvalues. Unless all the off-diagonal terms in $\cal
M$ are zero, one of the generically non-vanishing eigenvalues is positive
and the other is negative. As a result, there is always transient growth
in $P$. This is the reason for the peaks in Figures \ref{fg.simplepower},
\ref{fg.elnssimplepower} and \ref{fg.power}.

We complete the analysis of transient growth of $P$ for the ideal fluid
(see eqs.\ \ref{idealsound}). In this case we have
\beq
   {\cal M} = \left(\matrix{0 & -i kB\cr i kB & 0}\right),\qquad
   \qquad{\rm with\ eigenvalues}\qquad
   \lambda=\pm kB.
\label{idealtransient}\eeq
Transient growth can take place. We can parametrize all initial
conditions by an angle $\alpha$ and a phase $\beta$, by choosing
$\mathbf x=(\sin\alpha\exp[i\beta],\cos\alpha)$. The evolution of
the power spectrum of the energy obtained with specific initial
conditions from numerical solutions of eqs.\ (\ref{idealsound})
are shown in Figure \ref{fg.actualideal}.

For the evolution of the power spectrum of $\chi^1$ in a simple fluid, one has
\beq
   {\cal M} = \left(\matrix{-\frac83 & -4ik & 1\cr4 ik &0&0\cr1&0&0}\right),
      \qquad{\rm with\ eigenvalues}\qquad
   \lambda=0,\;\frac13\left[-4\pm\sqrt{25+144k^2}\right],
\label{simpletransient}\eeq
where we have taken $c_s^2=1/3$. Since one of the eigenvalues is positive,
transient growth occurs. Note that the mode $k=0$ can also display transient
growth. This analysis can be adapted to that for the Boltzmann fluid by
replacing the constants in $\cal M$ by appropriate functions of $\theta$.
However, the conclusions regarding transient growth carry over to that case.

For the power spectrum of $\chi^1$ in the ELNS approximation to the simple
fluid, one finds that
\beq
   {\cal M} = -\left(\matrix{
     2B & \frac43 i k\frac{1-c_s^2}{c_s^2}\cr 
     -\frac43 i k\frac{1-c_s^2}{c_s^2} & 0}\right),
      \qquad{\rm with\ eigenvalues}\qquad
   \lambda=B\pm\frac1{3c_s^2}\sqrt{9c_s^2B^2+16 k^2(1-c_s^2)^2}.
\label{elnssimpletransient}\eeq
Interestingly, the diffusive term, $c_s^2 k^2$, drops out of the transient
analysis for the energy density (it does appears in the transient analysis
for $y$). For $k=0$ transients do not grow. This seems to be the main
distinction between transient growth of energy density in ELNS and IS
dynamics.

The numerical results in Figure \ref{fg.actualelnssimplesound} seem
similar to the ideal fluid case at first sight. However, the diffusive
character of the underlying equations is manifested in two ways. First,
at fixed $k$, if one observes the time evolution, then one sees only a
single instance of transient growth, unlike the quasiperiodic behaviour
of $P_\epsilon$ in the ideal fluid. Second, at sufficiently large $k$
there is no transient growth, unlike the ideal fluid.

In the late time limit for a Boltzmann or conformal fluid one has, from
eq.\ (\ref{conformalintheta}),
\beq
     {\cal M} = \left(\matrix{
      2p & -\frac{ikBu_0}4{\rm e}^{p\theta} & \frac{u_0}4{\rm e}^{p\theta}\cr
      \frac{ikBu_0}4{\rm e}^{p\theta} & 0 & 0\cr
      \frac{u_0}4{\rm e}^{p\theta} & 0 & 0}\right),
     \quad{\rm with\ eigenvalues}\quad
     \lambda=0,p\pm\frac14\sqrt{16p^2+(1+B^2k^2)u_0^2\mathrm e^{2p\theta}}.
\label{conform}\eeq
Since one of the eigenvalues is positive, transient growth can take place.

\end{document}